\documentclass[10pt,aps,prb,twocolumn,superscriptaddress,showpacs,longbibliography,floatfix]{revtex4-2}

\usepackage{amssymb}
\usepackage{graphicx}
\usepackage{amsmath}
\usepackage{mathtools}
\usepackage[export]{adjustbox}
\usepackage{epsfig}
\usepackage{times}
\usepackage{color}
\usepackage{subfigure}
\usepackage{setspace}
\usepackage{natbib}
\usepackage[normalem]{ulem}
\usepackage{cancel}
\usepackage{soul}
\usepackage{physics}
\usepackage{multirow}

\usepackage[pdftex]{hyperref}
\usepackage{ulem} 
\hypersetup{colorlinks = true, urlcolor = blue, linkcolor = blue, citecolor = blue}

\usepackage[color=orange!60,textsize=footnotesize]{todonotes}
\usepackage{booktabs}

\AtBeginDocument{%
\heavyrulewidth=.08em
\lightrulewidth=.05em
\cmidrulewidth=.03em
\belowrulesep=.65ex
\belowbottomsep=0pt
\aboverulesep=.4ex
\abovetopsep=0pt
\cmidrulesep=\doublerulesep
\cmidrulekern=.5em
\defaultaddspace=.5em
}

\usepackage{xcolor}
\usepackage{xparse,xcoffins}
\ExplSyntaxOn
\NewCoffin\imagecoffin
\NewCoffin\labelcoffin
\keys_define:nn { miguel/label }
 {
  label   .tl_set:N = \l_miguel_label_tl,
  labelbox .bool_set:N = \l_miguel_label_box_bool,
  labelbox .default:n = true,
  fontsize .tl_set:N = \l_miguel_label_size_tl,
  fontsize .initial:n = \footnotesize,
  pos .choice:,
  pos/nw .code:n = \tl_set:Nn \l_miguel_label_pos_tl { left,up },
  pos/ne .code:n = \tl_set:Nn \l_miguel_label_pos_tl { right,up },
  pos/sw .code:n = \tl_set:Nn \l_miguel_label_pos_tl { left,down },
  pos/se .code:n = \tl_set:Nn \l_miguel_label_pos_tl { right,down },
  pos/n .code:n = \tl_set:Nn \l_miguel_label_pos_tl { hc,up },
  pos/w .code:n = \tl_set:Nn \l_miguel_label_pos_tl { left,vc },
  pos/s .code:n = \tl_set:Nn \l_miguel_label_pos_tl { hc,down },
  pos/e .code:n = \tl_set:Nn \l_miguel_label_pos_tl { right,vc },
  pos .initial:n = nw,
  unknown .code:n   = \clist_put_right:Nx \l_miguel_label_clist
                       { \l_keys_key_tl = \exp_not:n { #1 } }
 }
\clist_new:N \l_miguel_label_clist
\box_new:N \l_miguel_label_box
\box_new:N \l_miguel_label_image_box
\NewDocumentCommand{\xincludegraphics}{O{}m}
 {
  \group_begin:
  \tl_clear:N \l_miguel_label_tl
  \clist_clear:N \l_miguel_label_clist
  \keys_set:nn { miguel/label } { #1 }
  \tl_if_empty:NTF \l_miguel_label_tl
   {
    \miguel_includegraphics:Vn \l_miguel_label_clist { #2 }
   }
   {
    \SetHorizontalCoffin\imagecoffin
     {
      \miguel_includegraphics:Vn \l_miguel_label_clist { #2 }
     }
    \SetHorizontalCoffin\labelcoffin
     {
      \raisebox{\depth}
       {
        \bool_if:NTF \l_miguel_label_box_bool
         { \fcolorbox{white}{white}{\l_miguel_label_size_tl\l_miguel_label_tl} }
         { \l_miguel_label_size_tl\l_miguel_label_tl }
       }
     }
    \SetVerticalPole\imagecoffin{left}{3pt+\CoffinWidth\labelcoffin/2}
    \SetVerticalPole\imagecoffin{right}{\Width-3pt-\CoffinWidth\labelcoffin/2}
    \SetHorizontalPole\imagecoffin{up}{\Height-3pt-\CoffinHeight\labelcoffin/2}
    \SetHorizontalPole\imagecoffin{down}{3pt+\CoffinHeight\labelcoffin/2}
    \use:x{\JoinCoffins\imagecoffin[\l_miguel_label_pos_tl]\labelcoffin[vc,hc]} 
    \TypesetCoffin\imagecoffin
   }
   \group_end:
 }
\NewDocumentCommand{\setlabel}{m}
 {
  \keys_set:nn { miguel/label } { #1 }
 }
\cs_new_protected:Nn \miguel_includegraphics:nn
 {
  \includegraphics[#1]{#2}
 }
\cs_generate_variant:Nn \miguel_includegraphics:nn { V }
\ExplSyntaxOff

\begin{document}

\title{Compact Spin–Charge Separated Neural Quantum States for Valence-Bond States}


\author{Ang-Kun Wu}
\email{angkunwu@gmail.com}
\affiliation{Department of Physics and Astronomy, University of Tennessee, Knoxville, Knoxville, Tennessee, 37996, USA}
\author{Louis Primeau}
\affiliation{Department of Physics and Astronomy, University of Tennessee, Knoxville, Knoxville, Tennessee, 37996, USA}
\author{Yixin Zhang}
\affiliation{Department of Physics and Astronomy, University of Tennessee, Knoxville, Knoxville, Tennessee, 37996, USA}
\author{Jingtao Zhang}
\affiliation{Google, Mountain View, CA, 94043, USA}
\author{Adrian Del Maestro}
\affiliation{Department of Physics and Astronomy, University of Tennessee, Knoxville, Knoxville, Tennessee, 37996, USA}
\affiliation{Min H. Kao Department of Electrical Engineering and Computer Science, University of Tennessee, Knoxville, Tennessee 37996, USA}
\author{Yang Zhang}
\email{yangzhang@utk.edu}
\affiliation{Department of Physics and Astronomy, University of Tennessee, Knoxville, Knoxville, Tennessee, 37996, USA}
\affiliation{Min H. Kao Department of Electrical Engineering and Computer Science, University of Tennessee, Knoxville, Tennessee 37996, USA}

\date{\today}

\begin{abstract}
    Neural-network quantum states (NQS) provide a flexible nonlinear representation of quantum many-body wavefunctions, but their efficiency depends sensitively on whether the architecture reflects the sign structure and constrained Hilbert space of the target state. In this work, we propose a solvable-point-guided strategy: design the architecture at an exactly solvable point where the correct local rules can be read off, then refine to the non-exact regime by enlarging only the kernel size and hidden dimension.  The strategy is built from four physics-motivated designs: a stride-matched local-rule convolution, geometric pooling, a sign-resolving $\tanh(x^{2k+1})$ activation, and explicit spin-hole sector separation.  We test this approach on quasi-one-dimensional valence-bond-solid (VBS) states and their doped soliton variants (sVBS), the exact ground states of a $t$-$J$-like model with a single mobile hole.  In finite-size benchmarks, this architecture reaches high fidelity for the exact sVBS state with substantially fewer parameters than generic fully connected, convolutional, and transformer baselines tested under the same setup. For the spin sector, the learned local rule transfers from small to larger systems without retraining.   Away from the solvable point, increasing kernel size and hidden dimension systematically improves accuracy, and the model shows approximately $L^2$ parameter scaling in the gapless regime for system size $L$, compared with approximately $L^4$ for matrix-product states in the same regime.  Our work establishes a recipe for compact NQS in sign-structured, constrained Hilbert spaces and paves the pathway to physics-informed architectures for the broader $t$-$J$ and Hubbard families.
\end{abstract}

\maketitle


\section{Introduction}

Efficient representations of quantum many-body wavefunctions are central to modern condensed-matter physics. The Hilbert-space dimension of an interacting quantum system grows exponentially with system size, so even writing down a generic state quickly becomes impossible. A useful variational ansatz must therefore compress the wavefunction while retaining physically relevant structure, in particular its correlations, entanglement, and symmetries.

In one dimension, matrix-product states (MPS) provide a remarkably successful solution to this problem. They efficiently represent ground states of gapped local Hamiltonians and form the variational backbone of the density-matrix renormalization group (DMRG) \cite{White1992DMRG,Schollwock2011MPS,PerezGarcia2007MPS,CiracVerstraete2009,Orus2014}. From this perspective, the MPS may be viewed as a compression scheme organized directly around the entanglement structure of the state. At the same time, this strength also points to its limitation: the MPS is intrinsically a linear representation whose efficiency is controlled by the bond dimension. When the entanglement grows, as in critical or other area-law violating systems \cite{Tagliacozzo2008FiniteEntanglement,Pollmann2009FiniteEntanglement,Vidal2003Entanglement,PhysRevB.111.035119,Calabrese2004Entanglement,barghathi:2026rc}, doped systems \cite{StoudenmireWhite2012,LuoClark2019,Inui2021DeterminantFree,Roth2025HubbardSuperconductivity}, or higher-dimensional settings \cite{VerstraeteCirac2004PEPS,StoudenmireWhite2012,Orus2014}, the required bond dimension can increase rapidly, making the representation and optimization increasingly costly.

Building on the broader integration of machine-learning models into physics \cite{Carrasquilla2017,Mehta2019,Carleo2019ML,wu2025modeling}, neural-network quantum states (NQS) offer a complementary route to wavefunction representation \cite{Carleo2017NQS,GaoDuan2017,Deng2017,Nomura2017,Glasser2018,Sharir2020,HibatAllah2020,ChenHeyl2024,Rende2024LargeScaleNQS,Hul2026GrandCanonicalNQS}. Instead of organizing the state through a network of linear tensors, an NQS represents the amplitude of each many-body configuration through a nonlinear function of the microscopic degrees of freedom. This perspective suggests that neural networks may provide a more flexible compression of many-body states, especially when physically relevant features are not efficiently captured by a strictly linear ansatz. Indeed, NQS have attracted broad interest as variational wavefunctions for spin systems \cite{Nomura2017,Sharir2020,HibatAllah2020,ChenHeyl2024}, fermionic systems \cite{LuoClark2019,Inui2021DeterminantFree,Hermann2020,Pfau2020,rc31-5hl9}, and frustrated models \cite{Glasser2018,ChenHeyl2024,Rende2024LargeScaleNQS,Machaczek2025FractonNQS,Gu2026ParetoBackflow,ybgv-35jm}, precisely because they combine expressive nonlinear parametrizations with scalable gradient-based optimization.

However, the success of an NQS is not determined by expressive power alone. Generic black-box architectures can miss problem-specific symmetries and local constraints, often degrading representational efficiency and training quality \cite{Choo2018SymmetriesExcitedNQS,Nomura2021RestoringSymmetry,ChenHeyl2024,Rende2024LargeScaleNQS}. More broadly, variational optimization can suffer from severe trainability pathologies in poorly chosen ansatz/cost settings, including barren-plateau phenomena established for parameterized quantum circuits \cite{McClean2018BarrenPlateaus,Cerezo2021CostFunctionBP,Larocca2025BarrenPlateausReview}. A natural question is therefore how to design neural-network architectures that respect the sign structure and constrained Hilbert space of strongly correlated states, in particular the doped Mott systems described by the $t$-$J$ and Hubbard models, rather than being chosen only by analogy with standard machine-learning tasks \cite{Chen2018,Glasser2018,LuoHalverson2023InfiniteNQS,ChenHeyl2024,Rende2024LargeScaleNQS,Cortes2026BasisDependence}. Exactly solvable many-body states provide a useful template for this question: their local rules, symmetries, and factorization properties can be read off directly, and the resulting architecture can then be refined to nearby non-exact regimes \cite{Carleo2017NQS}.

In this work, we propose a solvable-point-guided
strategy for NQS design and demonstrate it on quasi-one-dimensional valence-bond-solid (VBS) states \cite{Anderson1973RVB} and their doped soliton variants (sVBS) \cite{Affleck1987AKLT,PhysRevB.107.L140401,glittum2025resonant}, the exact ground states of a $t$-$J$-like model with a single mobile hole. We identify four physics-motivated designs: a stride-matched local-rule convolution, geometric pooling, a sign-resolving $\tanh(x^{2k+1})$ activation, and explicit spin-hole sector separation that encodes the fermion sign structure of the doped state. This spin-hole separation is the architectural analogue of the spin-charge separation characteristic of doped one-dimensional systems. Together these designs form a minimal architecture that captures both the local singlet pattern and the global resonant motion of the hole.

At the exactly solvable point, this architecture reproduces the sVBS ground state exactly. Away from this point, we enlarge the kernel size and hidden dimension to absorb the additional entanglement, and the parameter count grows as $L^2$ with system size $L$. In contrast, the MPS bond dimension grows as $L^4$ in the gapless regime. For the spin-only VBS sector, the learned representation is also size-independent: a network trained on $L=2$--$10$ transfers directly to $L=12$--$20$ at the exactly solvable point, showing that the network has learned a size-independent local rule rather than a finite-size fit.

The same designs apply beyond the sVBS test bed. The spin-hole sector separation directly encodes the fermion sign structure of doped Mott states, which is also the origin of the sign problem in the broader $t$-$J$ and Hubbard families. Our solvable-point-guided strategy therefore provides a constructive route from exactly solvable wavefunctions to physics-informed NQS for strongly correlated electron systems.

The manuscript is organized as follows. In Sec. \ref{sec:2}, we introduce the minimal neural-network representation of the quasi-one-dimensional VBS state and identify the local pairing rule, geometric pooling, and activation structure required for an exact representation of the VBS state. In Sec. \ref{sec:3}, we generalize this construction to soliton VBS states with one hole and show how the architecture captures both the singlet background and the resonant hole motion. In Sec. \ref{sec:4}, we study the regime away from the exactly solvable point and show that the same VBS-inspired architecture remains effective as a systematically improvable ansatz. In Sec. \ref{sec:5}, we discuss energy-based optimization of the NQS using variational Monte Carlo. In Sec. \ref{sec:6}, we analyze translational structure and system-size extrapolation. Finally, in Sec. \ref{sec:7}, we summarize the main conclusions and discuss implications for more general neural-network representations of quantum many-body states.

\section{Minimal Neural-Network Representation of the VBS State}\label{sec:2}

We begin with the quasi-one-dimensional VBS state, which is a product of nearest-neighbor spin singlets. For a chain of even length $L$,
\begin{equation}
    \begin{split}
    \ket{\mathrm{VBS}}
    &= \prod_{j=1}^{L/2} \frac{1}{\sqrt{2}}
    \left(\ket{\uparrow\downarrow}-\ket{\downarrow\uparrow}\right)_{2j-1,2j} \\
    &= \left(\frac{1}{\sqrt{2}}\right)^{L/2}
    \sum_{s=\{\sigma_j\}} A_s \ket{s},
    \end{split}
\end{equation}
where $A_s\in\{-1,0,1\}$ denotes the unnormalized amplitude in the spin basis $\ket{s}$ \cite{Affleck1987AKLT} where $\sigma_j =\, \uparrow,\downarrow$.
Although this state is simple in its singlet-product form, representing it directly in the many-body basis still requires learning a nontrivial sign and selection rule. We denote by $\Psi_\theta(s)$
the neural-network amplitude for basis configuration $s$. For small systems, where the Hilbert space can be fully enumerated, we train in a supervised manner by fitting $\Psi_\theta(s)$ to the exact VBS amplitudes $A_s$ with mean-squared loss,
$\mathcal{L}(\theta)=\frac{1}{D_H}\sum_{s\in\mathcal D_H}\left|\Psi_\theta(s)-2^{-L/4}A_s\right|^2,$ where $\mathcal D_H$ is the full Hilbert space of dimension $D_H$. We use this direct-amplitude objective rather than a log-amplitude target $\log(A_s)$, because $A_s$ can take zero and negative values.

\begin{figure}[t!]
\begin{center}
\includegraphics[width=1.0\linewidth]{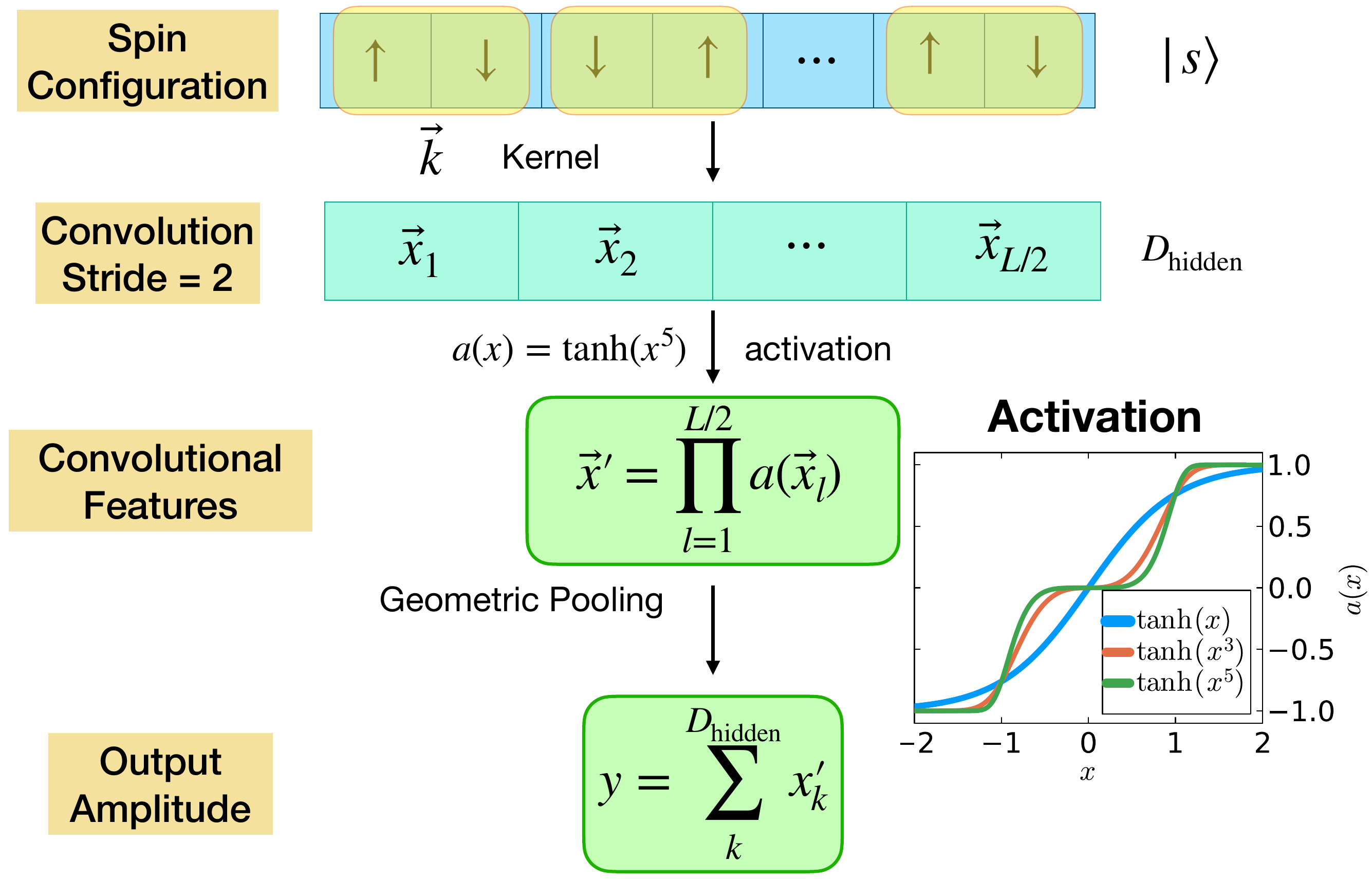}
\caption{\label{fig:PhaseNN} 
    The neural-network architecture for the quasi-one-dimensional VBS state. A convolutional layer with kernel size $2$ and stride $2$ extracts the local singlet-pairing rule on each dimer. The activation function $\tanh(x^{2k+1})$ separates the three local classes with amplitudes $+1$, $0$, and $-1$. Geometric pooling combines the resulting features multiplicatively, reflecting the product structure of the VBS wavefunction. A final fully connected layer outputs the scalar wavefunction coefficient. Inset: $\tanh(x^{2k+1})$ for different values of $k$. 
    Input: each spin is encoded as a one-hot vector, $\uparrow \to (1,0)^T, \downarrow \to (0,1)^T$ (expanded to $4$ dimensional space for zero and double occupancy), processed as input channels for convolution.
}
\end{center}
\end{figure}

The key observation is that the amplitude factorizes into independent local rules on each dimer. For every nearest-neighbor pair, an antiparallel configuration contributes a nonzero weight, while a parallel configuration gives zero:
\begin{equation}\label{eq:spins}
    \ket{\uparrow\downarrow} \mapsto +1,\qquad
    \ket{\downarrow\uparrow} \mapsto -1,\qquad
    \ket{\uparrow\uparrow},\,\ket{\downarrow\downarrow} \mapsto 0.
\end{equation}
Therefore, the full coefficient $A_s$ is simply the product of these local dimer contributions. This factorized structure immediately suggests a minimal neural-network architecture. A one-dimensional convolution layer with kernel size $2$ and stride $2$ extracts the local singlet-pairing features on each dimer, as illustrated in Fig. \ref{fig:PhaseNN}. The resulting local features are then combined by a geometric pooling layer, which multiplies rather than averages them. This choice is essential: unlike max or average pooling, geometric pooling matches the multiplicative structure of the VBS amplitude and naturally incorporates the factor $(1/\sqrt{2})^{L/2}$.

The local dimer rule is effectively a three-class classification problem with outputs $\{+1,0,-1\}$. To separate these classes, especially the zero-amplitude sector, we introduce the activation function
\begin{equation}
    a(x)=\tanh(x^{2k+1}),\qquad k=1,2,\cdots.
\end{equation}
Compared with the standard $\tanh(x)$, the additional odd power broadens the flat region near the origin and improves the separation between the $0$ class and the $\pm1$ classes, as seen in the inset of Fig.~\ref{fig:PhaseNN}. In practice, this modification is important for accurately learning the local VBS rule. 
Although the derivative of the activation vanishes at $x=0$, which can disrupt optimization during random initialization \cite{Glorot2010,He2015,Pascanu2012,Singh2020}, this does not present a practical obstacle in our training process. The vanishing gradient occurs only at the isolated saddle point $x=0$, where the second derivative also vanishes; more importantly, the flat curvature near the origin broadens the transition region and sharpens the decision boundaries between the three-valued local rule $\{+1, 0, -1\}$ and enables $\Psi_\theta(s)$ to fit the exact VBS amplitudes with high accuracy via mean-squared loss.

\begin{figure}[b!]
\begin{center}
\includegraphics[width=0.99\linewidth]{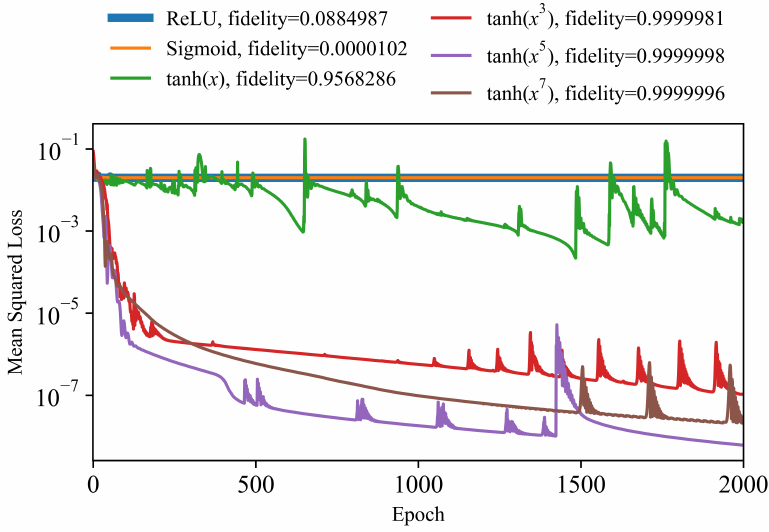}
\caption{\label{fig:activation} 
    Mean squared loss (MSL) for learning the local pairing rule of the VBS state with different activation functions at system size $L=16$, with Hilbert-space dimension $12{,}870$. In all cases, the neural network uses kernel size $2$ and hidden dimension $16$ with learning rate $10^{-2}$. 
}
\end{center}
\end{figure}

The numerical comparison in Table~\ref{tab:localrule} confirms this behavior. Relative to $\tanh(x)$, the activations $\tanh(x^3)$ and $\tanh(x^5)$ substantially reduce the error for the $0$ class while maintaining high accuracy for the $\pm1$ classes. Figure~\ref{fig:activation} shows the same trend throughout training: $\tanh(x^{2k+1})$ consistently outperforms $\tanh(x)$, while $\mathrm{ReLU}(x)$ and $\mathrm{sigmoid}(x)$ do not show learning over the range of epochs considered and lead to significantly larger errors. Empirically, $k=2$ gives slightly better performance than $k=1$, whereas larger $k$ does not lead to a clear additional gain.

\begin{table}[t!]
    \centering
    \renewcommand{\arraystretch}{1.5}
    \setlength\tabcolsep{8pt}
    \begin{tabular}{@{}lccc@{}} 
        \toprule
         & $\mathbf{\vb*{L}=2-10}$ & \multicolumn{2}{c}{$\mathbf{\vb*{L}=20}$} \\ 
        \textbf{activation} & \textbf{training} & \textbf{1,-1} & \textbf{0} \\
        \midrule
        $\tanh(x)$ & $2.7\times 10^{-7}$ & $1.7\times 10^{-6}$ & $1.8\times 10^{-2}$ \\
        $\tanh(x^3)$ & $1.0\times 10^{-7}$ & $7.7\times 10^{-8}$ & $2.8\times 10^{-4}$ \\
        $\tanh(x^5)$ & $4.6\times 10^{-9}$ & $8.2\times 10^{-8}$ & $8.9\times 10^{-5}$ \\
        \bottomrule
    \end{tabular}
    \caption{Mean squared loss (MSL) for learning the local pairing rule of the VBS state in a single run. The neural network has kernel size $2$, hidden dimension $16$ after the convolutional layer, and $161$ trainable parameters. It is trained on system sizes $L=2$--$10$ and tested at $L=20$. 
    The MSL is reported separately for configurations with antiparallel dimer pairs ($\uparrow\downarrow$ or $\downarrow\uparrow$), which contribute amplitudes $\pm 1$, and configurations containing at least one parallel dimer pair ($\uparrow\uparrow$ or $\downarrow\downarrow$), which contribute amplitude $0$, according to Eq. \ref{eq:spins}.
    }
    \label{tab:localrule}
\end{table}

In the pure VBS limit, the architecture can in principle be reduced to a single convolution layer with kernel size $2$ and hidden dimension $1$. In practice, however, such a minimal network is harder to optimize and is more susceptible to trapping in local minima. We find that using hidden dimension $\gtrsim 8$ leads to substantially more stable training, while keeping the total number of parameters small.

As a baseline, it is also instructive to compare this physics-informed architecture with a generic fully connected neural network, where we flatten the one-hot-encoded input. A fully connected model with three hidden layers of width $1024$ reaches a mean squared loss of order $10^{-4}$ for the $L=14$ VBS state, despite having roughly $1.1\times10^6$ parameters. Thus, a naive fully connected neural network does not appear to systematically capture the entanglement structure of the state \cite{Paul2026EntanglementBound} even though the entanglement entropy and bond dimension are low. In contrast, the convolutional architecture proposed here achieves far higher accuracy with only $161$ parameters, independent of system size. This comparison highlights the main point of this section: when the neural-network design is matched to the local structure of the quantum state, the representation becomes both more compact and more accurate.

\section{Soliton VBS States: Exact Structure and Neural-Network Representation}\label{sec:3}

The minimal neural-network representation of a pure VBS state provides a natural starting point for constructing neural-network quantum states for a perturbed state containing a single mobile soliton. We consider quasi-one-dimensional states formed by superposing VBS configurations with a single hole at different positions. The corresponding sVBS state can be written as \cite{PhysRevB.107.L140401}
\begin{equation}\label{eq:sVBS}
    \begin{split}
        \ket{\Psi_\mathrm{sVBS}} &= \sum_{j=1}^L a_j \ket{j,\mathrm{VBS}},\\
        \ket{j,\mathrm{VBS}} &= \left(\prod_{k=1}^{(L-1)/2} c^\dagger_{x_{2k-1},x_{2k}} \right) \ket{\Omega},\\
        c^\dagger_{i,j} &\equiv \frac{1}{\sqrt{2}}
    \left(c^\dagger_{i\uparrow}c^\dagger_{j\downarrow}-c^\dagger_{i\downarrow}c^\dagger_{j\uparrow}\right)
    \end{split}
\end{equation}
where $\ket{j,\mathrm{VBS}}$ denotes the VBS configuration with a hole at site $j$, and $a_j$ depends only on the hole position. Here $x_k$ labels the occupied sites after 
creating a hole at site $j$, and $c^\dagger_{i,j}$ creates a singlet pair between sites $i$ and $j$.
The hole therefore acts as a soliton separating two dimerization patterns.

A useful feature of this construction is that the local singlet-pairing rule remains the same as in the pure VBS state and is independent of system size. In principle, one may therefore pretrain the convolutional part of the network on the local spin structure and freeze it when moving to larger systems, leaving only the hole wavefunction to be learned. This effectively reduces the problem to learning the single-particle motion of the hole in an emergent tight-binding background. Such pretraining can reduce model complexity by the convolution layer and improve optimization slightly for medium-sized $L\sim 21$ systems under variational Monte Carlo (Appendix \ref{app:vmc-bench}). However, for large systems, this approach does not scale as the fully connected network must fit an effective tight-binding wavefunction whose finite-size excitation gap closes with $L$, so many low-energy states become nearly degenerate and optimization can no longer reliably isolate the true ground state. Nevertheless, our broader goal is to construct an NQS architecture that can be trained end to end, without imposing this separation by hand.

\begin{figure}[t!]
\begin{center}
\includegraphics[width=0.95\linewidth]{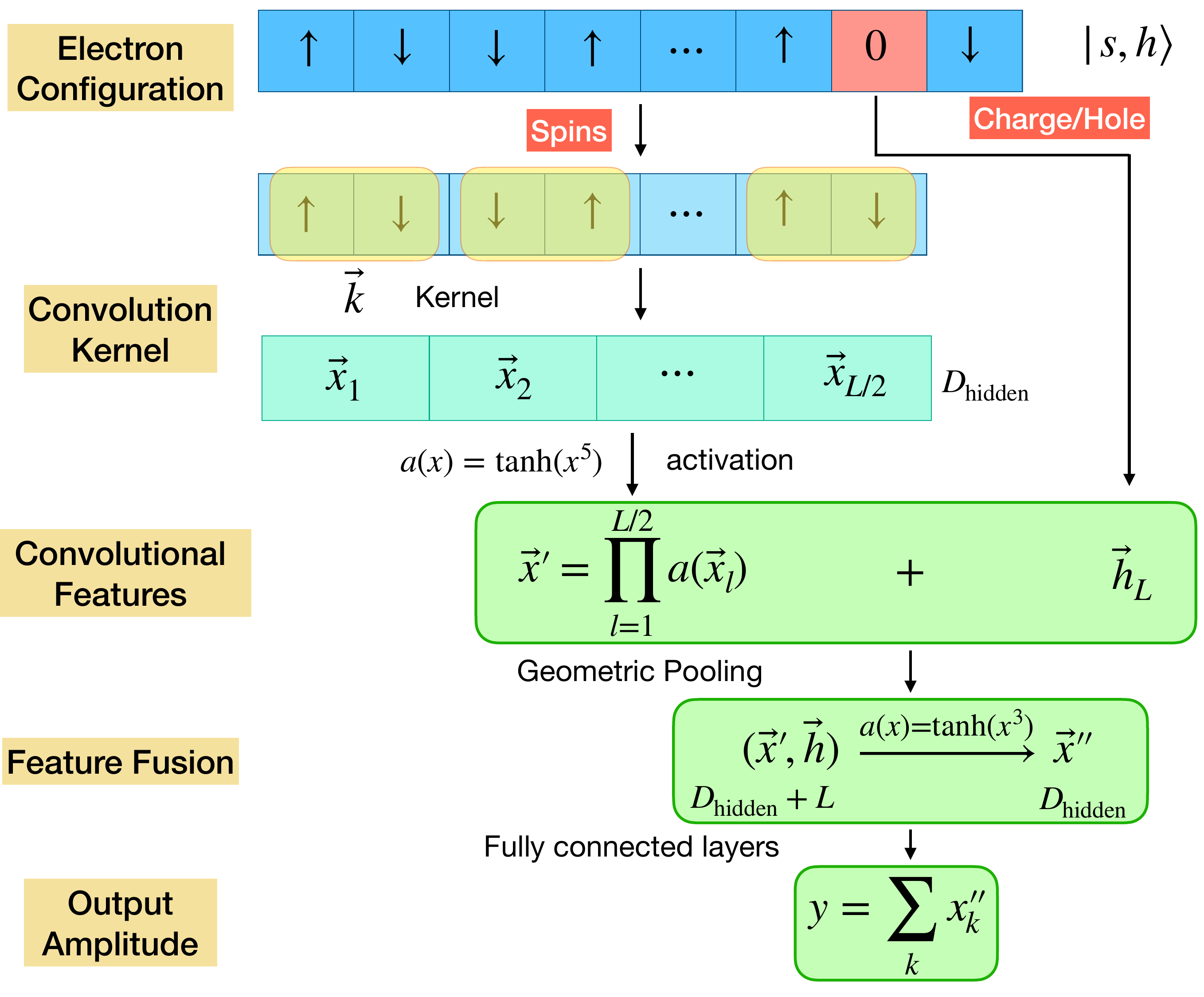}
\caption{\label{fig:fullNN} 
Neural-network architecture for the quasi-one-dimensional soliton VBS state with a single hole. The input configuration is separated into spin and hole degrees of freedom. The spin branch follows the same structure as in Fig. \ref{fig:PhaseNN}, up to the final output layer. After geometric pooling, the spin features are combined with the hole-position features and passed through a small fully connected layer to capture the resonant superposition of different VBS configurations. The final activation function is $\tanh(x^3)$, which empirically performs better than $\tanh(x)$ and $\mathrm{ReLU}(x)$.
}
\end{center}
\end{figure}

The corresponding neural-network ansatz is shown in Fig.~\ref{fig:fullNN}. 
In contrast to the pure-spin VBS network in Fig.~\ref{fig:PhaseNN}, the input is decomposed into spin and hole sectors, which are processed separately and recombined only at the end. The convolutional layer with kernel size $2$ and stride $2$ continues to encode the local singlet structure, while the fully connected layer captures the resonant superposition over different hole positions. Geometric pooling is retained to represent the product structure of the singlet pairs, and the activation function $\tanh(x^{2k+1})$ is inherited from the pure VBS case to resolve the local pairing rule.

\begin{table}[t]
    \centering
    \renewcommand{\arraystretch}{1.5}
    \setlength\tabcolsep{8pt}
    \begin{tabular}{@{}crc@{}} 
        \toprule
        $\vb*{L}$ &  $\vb*{D_H}$ & $\vb*{1-|\bra{\psi_\mathrm{exact}}\ket{\psi_\mathrm{NN}}|}$ \\
        \midrule
        13 & $12{,}012$ & $2.38\times 10^{-7}$ \\
        15 & $51{,}480$ & $5.98\times 10^{-7}$ \\
        17 & $218{,}790$ & $2.98\times 10^{-6}$ \\
        \bottomrule
    \end{tabular}
    \caption{\label{tab:sl_exact}
    Supervised learning of the exact many-body sVBS state for $t_2=0.5t_1$ and $J_1=J_2=0$. The neural network uses kernel size $2$ and hidden dimension $16$, and is trained with squared loss on the normalized wavefunction for $2000$ epochs. The fidelity is evaluated in the full Hilbert space (dimension $D_H$) after normalization and taken as the highest fidelity over $10$ independent runs.
    }
\end{table}

The sVBS states arise as exact ground states of a $t$-$J$-like model with a single hole, which may be viewed as a doped counterpart of the pure VBS system \cite{PhysRevB.107.L140401}. The Hamiltonian is given by
\begin{equation}\label{eq:Ham}
\begin{split}
    H &= -t_1\sum_{\langle i,j\rangle,\sigma}
    \left[c^\dagger_{i,\sigma}c_{j,\sigma} +\mathrm{H.c.}\right]\\
    &\quad -t_2\sum_{\langle\langle i,j\rangle\rangle,\sigma,\; i,j\,\mathrm{odd}}
    \left[c^\dagger_{i,\sigma}c_{j,\sigma} +\mathrm{H.c.}\right]\\
    &\quad +J_1 \sum_{\langle i,j\rangle} \mathbf{S}_i\cdot \mathbf{S}_j
    +J_2 \sum_{\langle \langle i,j\rangle\rangle,\; i,j\, \mathrm{odd}} \mathbf{S}_i\cdot \mathbf{S}_j,
\end{split}
\end{equation}
with
\begin{equation}
    \mathbf{S}_j=(S_j^x,S_j^y,S_j^z),\qquad
    \mathbf{S}_j=\frac{1}{2}\sum_{\sigma\sigma'}
    c^\dagger_{j\sigma}\boldsymbol{\sigma}_{\sigma\sigma'}c_{j\sigma'},
\end{equation}
where 
$\sigma,\sigma'\in\{\uparrow,\downarrow\}$ are spin indices and $\boldsymbol{\sigma}=(\sigma_x,\sigma_y,\sigma_z)$ is a vector of Pauli matrices (acting on the spin indices). The hole hops to nearest-neighbor sites with amplitude $t_1$ and to 
next-nearest-neighbor sites (only among odd-indexed sites) with amplitude $t_2$. For $t_1,t_2>0$ and $J_1=J_2\ge 0$, this model has an exact sVBS ground state. As usual in the $t$-$J$ setting, double occupancy is excluded by an infinite on-site repulsion. 

Because the Hamiltonian in Eq.~\ref{eq:Ham} conserves both the total particle number and $S^z$, its action on a basis state $\ket{j,\mathrm{VBS}}$ moves the hole to neighboring allowed positions, provided the spin configuration is compatible with the hopping process. This observation already suggests that the many-body problem reduces to the motion of a hole on top of a rigid singlet background.

The exactness of the sVBS state for $t_1,t_2>0$ and $J_1=J_2$ can be understood via a local decomposition.  The above quasi-one-dimensional Hamiltonian is the so-called ``sawtooth" or ``delta" $\Delta$ chain due to the geometry of the lattice \cite{Rausch2023QuantumSpinSpiral,PhysRevB.107.L140401,Jiang2015Sawtooth,Yamaguchi2020AsymmetricJ1J2,Monti1991DeltaTrees,RuizPerez2000Malonate,TonegawaKaburagi2004DeltaChain,Kaburagi2005AnisotropicDelta,Inagaki2005DeltaChainHF,Blundell2003QuantumTopological}. The Hamiltonian can be decomposed into triangular $\Delta$ 
terms,
\begin{equation}
    H=\sum_{\alpha=1}^{(L-1)/2} h_\alpha,
\end{equation}
where $h_\alpha$ acts on sites $2\alpha-1$, $2\alpha$, and $2\alpha+1$ and $\alpha$ is the triangle index:
\begin{equation}
\begin{split}
    h_\alpha &= h_\alpha^t + h_\alpha^J,\\
    h_\alpha^t &= -t_1\sum_{\sigma}
    \left[c^\dagger_{2\alpha-1,\sigma}c_{2\alpha,\sigma}
    +c^\dagger_{2\alpha,\sigma}c_{2\alpha+1,\sigma}
    +\mathrm{H.c.}\right]\\
    &\quad -t_2\sum_{\sigma}
    \left[c^\dagger_{2\alpha-1,\sigma}c_{2\alpha+1,\sigma}
    +\mathrm{H.c.}\right],\\
    h_\alpha^J &= J_1 \mathbf{S}_{2\alpha-1}\cdot \mathbf{S}_{2\alpha}
    +J_1 \mathbf{S}_{2\alpha}\cdot \mathbf{S}_{2\alpha+1}
    +J_2 \mathbf{S}_{2\alpha-1}\cdot \mathbf{S}_{2\alpha+1}.
\end{split}
\end{equation}
It is therefore sufficient to analyze one triangle: each $h_{\alpha}$ is local to three sites. Below we show that, for $t_1,t_2>0$ and $J_1=J_2$, a state with one singlet pair on each triangle minimizes both the exchange and hopping terms. This local result implies that the sVBS manifold is invariant under $H$, the hole moves as a free particle in the singlet background, and the sVBS state is the exact ground state.

When $J_1=J_2$, the spin part can be simplified as
\begin{equation}
    h_\alpha^J
    =\frac{J_1}{2}
    \left(\mathbf{S}_{\mathrm{tot}}^2-\mathbf{S}_{2\alpha-1}^2-\mathbf{S}_{2\alpha}^2-\mathbf{S}_{2\alpha+1}^2\right),
\end{equation}
where $\mathbf{S}_{\mathrm{tot}}=\mathbf{S}_{2\alpha-1}+\mathbf{S}_{2\alpha}+\mathbf{S}_{2\alpha+1}$.  Since $\mathbf{S}_j^2=3/4$ for spin-$1/2$ and the minimum allowed value is $\mathbf{S}_{\mathrm{tot}}^2=3/4$ for three spins and $\mathbf{S}_{\mathrm{tot}}^2=0$ for two spins, the minimum eigenvalue of $h_\alpha^J$ is $-3J_1/4$. This minimum is realized by a singlet pair within the triangle. Hence any configuration with one singlet on each triangle, namely a VBS configuration with one hole, already minimizes the spin part of the Hamiltonian.

For the hopping term $h_\alpha^t$, the one-hole, two-spin sector on each triangle decomposes into singlet and triplet channels; diagonalizing the resulting $3\times 3$ matrices (Appendix~\ref{app:hopping}) shows that the singlet channel has the lower minimum eigenvalue for $t_1,t_2>0$, so both the exchange and hopping parts of $h_\alpha$ select the same local structure: one hole moving in a singlet background.

This immediately explains why the sVBS manifold is closed under the action of the Hamiltonian. Starting from a basis state $\ket{j,\mathrm{VBS}}$, the exchange terms only measure the local singlet structure, while the hopping terms move the hole to another allowed position without destroying the surrounding singlet pattern. The Hamiltonian therefore does not generate states outside the subspace spanned by $\{\ket{j,\mathrm{VBS}}\}_{j=1}^L$ and thus $\ket{\Psi_\mathrm{sVBS}}$ is a candidate for the exact ground state, with coefficients $a_j$ determined later in this section.

The exact state also admits a simple entanglement characterization. Whereas the pure VBS state has maximum matrix-product-state bond dimension $\chi=2$, the sVBS state with one hole has maximum bond dimension $\chi=3$, reflecting the coexistence of the singlet background and the mobile defect. For comparison, a generic single-particle wavefunction, $\ket{\psi} = \sum_{j=1}^L a_j \ket{j}$, 
has bond dimension $\chi=2$ across a cut at position $k$, which separates the left and right parts of the 1D chain, since
\begin{equation}
    \begin{split}
    \ket{\psi}
    &= \sum_{j=1}^k a_j \ket{j}_L \otimes \ket{0}_R
    +\sum_{j>k}^L a_j \ket{0}_L\otimes \ket{j}_R\\
    &= \ket{\psi}_L\otimes\ket{0}_R
    +\ket{0}_L\otimes \ket{\psi}_R.
    \end{split}
\end{equation}
Here $\ket{j}_L$ ($\ket{j}_R$) denotes the state with an electron at site $j$ on the left (right) side of the cut, $c_j^\dagger \ket{0}, j\in L (R)$, and $\ket{0}_L$ ($\ket{0}_R$) denotes the vacuum state on the left (right) side.
For the sVBS state, the structure depends on whether the cut breaks a singlet. If $k$ is even and the hole lies to the left of the cut, one singlet is broken; if the hole lies to the right, the state remains factorized:
\begin{equation}
    \begin{split}
        \ket{\Psi_\mathrm{sVBS}}
        &= \sum_{j=1}^k a_j\bigg[
        \ket{\bar{j}}_L\otimes\frac{1}{\sqrt{2}}\ket{\uparrow}_L\otimes\ket{\downarrow}_R\otimes\ket{\bar{0}}_R\\
        &\qquad\qquad
        -\ket{\bar{j}}_L\otimes\frac{1}{\sqrt{2}}\ket{\downarrow}_L\otimes\ket{\uparrow}_R\otimes\ket{\bar{0}}_R
        \bigg]\\
        &\quad +\sum_{j>k}^L a_j \ket{0}_L\otimes \ket{j}_R\\
        &= \ket{\bar{\psi}\uparrow}_L \otimes \ket{\downarrow\bar{0}}_R
        -\ket{\bar{\psi}\downarrow}_L \otimes \ket{\uparrow\bar{0}}_R
        \\
        &+\ket{0}_L\otimes \ket{\psi}_R,
    \end{split}
\end{equation}
where $\ket{\bar{j}}$ denotes the state with the broken-pair site removed, $\ket{\bar{0}}$ denotes the remaining paired background, $\ket{\bar{\psi}\sigma}_L
\equiv \sum_{j=1}^k \frac{a_j}{\sqrt{2}}\ket{\bar{j}}_L\otimes\ket{\sigma}_L$, $\ket{\sigma\bar{0}}_R
\equiv \ket{\sigma}_R\otimes \ket{\bar{0}}_R$, and $\ket{\psi}_R\equiv \sum_{j>k}^L a_j \ket{j}_R.$
This decomposition shows that $\chi=3$.

If instead $k$ is odd, the roles are reversed: configurations with $j\le k$ do not break a singlet, whereas those with $j>k$ do. Similarly, one finds
\begin{equation}
    \begin{split}
        \ket{\Psi_\mathrm{sVBS}}
          &= \ket{\psi}_L\otimes \ket{0}_R
          +\ket{\bar{0}\uparrow}_L\otimes\ket{\downarrow\bar{\psi}}_R
        \\
          &-\ket{\bar{0}\downarrow}_L\otimes \ket{\uparrow\bar{\psi}}_R,
    \end{split}
\end{equation} 
%
%
Hence $\chi=3$ for odd cuts as well. We therefore conclude that the exact sVBS ground state has maximum bond dimension $\chi=3$, which can be verified by running DMRG (ITensor.jl \cite{ITensor}) in the large sweep limit \cite{White1992DMRG,Schollwock2011MPS}. 

This structure also clarifies the effective low-energy description. Once the singlet background is fixed, the Hamiltonian restricted to the subspace spanned by $\ket{j,\mathrm{VBS}}$ reduces to a tight-binding problem for the hole coordinate. The resulting effective Hamiltonian in the basis $\{\ket{j,\mathrm{VBS}}\}_{j=1}^L$ is
\begin{equation}\label{eq:eff}
\begin{split}
    H_L &=
    \begin{pmatrix}
        0 & -t_1 & 0 & \cdots & 0 & 0\\
        -t_1 & 0 & -t_1 & \cdots & 0 & 0\\
        \vdots & \vdots & \vdots & \cdots & \vdots & \vdots\\
        0 & 0 & 0 & \cdots & -t_1 & 0
    \end{pmatrix}\\
    &\quad +
    \begin{pmatrix}
        0 & 0 & -t_2 & \cdots & 0 & 0 & 0\\
        0 & 0 & 0 & \cdots & 0 & 0 & 0\\
        \vdots & \vdots & \vdots & \cdots & \vdots & \vdots & \vdots\\
        0 & 0 & 0 & \cdots & -t_2 & 0 & 0
    \end{pmatrix},
\end{split}
\end{equation}
which has the same quadratic form as the underlying noninteracting hopping problem. The essential point is that the allowed hoppings move the hole without destroying the local singlet structure. The ground state of this effective Hamiltonian gives the coefficients $a_j$ in the sVBS state in Eq.~\eqref{eq:sVBS}, which are determined by the resonant motion of the hole in the singlet background.

\begin{table}[t!]
    \centering
    \renewcommand{\arraystretch}{1.5}
    \setlength\tabcolsep{6pt}
    \begin{tabular}{@{}rrll@{}} 
        \toprule
        NN architecture & Parameters & Loss (MSL) & Fidelity \\ [0.5ex]
        \midrule
        FCNN & $1{,}104{,}897$ & $7.946\times 10^{-5}$ & $0.2511731$ \\
        Conv1D & $25{,}601$ & $8.325\times 10^{-5}$ & $0.0$ \\
        Transformer & $199{,}361$ & $8.324\times 10^{-5}$ & $0.0048141$ \\
        VBS NN & $641$ & $2.646\times 10^{-9}$ & $0.9999867$ \\
        \bottomrule
    \end{tabular}
    \caption{
    Supervised learning of the exact many-body sVBS state for $L=13$, $t_2=0.5t_1$, and $J_1=J_2=0$. The fully connected neural network (FCNN) has three hidden layers of width $1024$. The convolutional network (Conv1D) has three convolutional layers with kernel size $3$ and hidden dimension $64$; it collapses to an almost-zero prediction because the output is averaged over incompatible local structures. The transformer uses a standard vision-transformer-style architecture with patch size $2$ for pairing \cite{Dosovitskiy2021ViT,Yamazaki2026TransformerNQS}. Our VBS-inspired network uses kernel size $2$ and hidden dimension $16$, and is trained for $2,000$ epochs with Adam.
    }\label{tab:architectures}
\end{table}

Although this state is a superposition of only $L$ VBS configurations, it is represented in a much larger many-body Hilbert space of dimension
\begin{equation}
    D_H=L\binom{L-1}{(L-1)/2},
\end{equation}
which grows exponentially with system size.
This makes the many-body representation problem nontrivial: the neural network must simultaneously encode the local singlet rule and the global resonant superposition over hole positions. As shown in Table~\ref{tab:architectures}, generic architectures such as fully connected networks, naive convolutional networks, and transformers perform poorly on this task. In particular, architectures that do not explicitly separate the spin and hole degrees of freedom tend to average over incompatible local structures and fail to reproduce the exact state. By contrast, the VBS-inspired architecture proposed here achieves near-unit fidelity with a very small number of parameters. The central design principle is therefore clear: by isolating the spin and hole sectors and combining them only after the local singlet structure has been extracted, the network can capture both the local pairing physics and the resonant motion of the soliton.

\section{Beyond the Exactly Solvable Point}\label{sec:4}

\begin{figure}[t!]
\begin{center}
\includegraphics[width=0.95\linewidth]{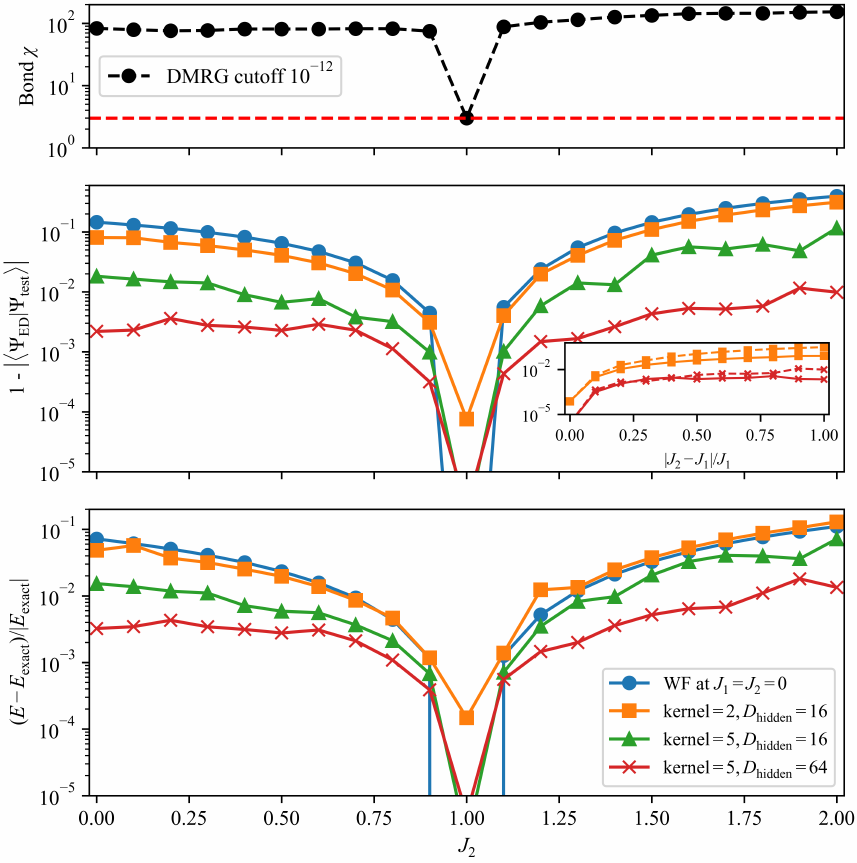}
\caption{\label{fig:jscompare}
    Comparison of ground states as a function of $J_2\in[0,2]$ at fixed $t_1=1$, $t_2=0.5$, and $J_1=1$ for system size $L=13$. The top, middle, and bottom panels show the MPS bond dimension $\chi$, the neural-network fidelity, and the relative energy error, respectively. At the exactly solvable point $J_2=J_1$, the ground state is the sVBS state and the minimal network with kernel size $2$ and hidden dimension $16$ is sufficient to achieve a relative infidelity of $\sim 10^{-4}$. Away from this point, the fidelity to the exact sVBS state decreases, while increasing the kernel size and hidden dimension systematically improves the neural-network approximation. From the DMRG perspective, the departure from the exact point is marked by a sudden increase of the required bond dimension above $\chi=3$. 
    Inset: fidelity of the two networks versus $|J_2-J_1|/J_1$ (solid: $J_2<J_1$, dashed: $J_2>J_1$).
}
\end{center}
\end{figure}

Although the sVBS state ceases to be the exact ground state once the couplings move away from the solvable point, the same neural-network architecture remains a highly effective variational ansatz. When $J_1=J_2$, the sVBS state is the exact ground state; when $J_1\neq J_2$, it is not. This change is reflected most clearly in the entanglement structure: from the DMRG point of view, the bond dimension increases abruptly once the system leaves the exact point, indicating additional correlations beyond those captured by the minimal sVBS construction.

Nevertheless, the VBS-inspired neural-network architecture continues to provide a useful starting point even when $J_1 \ne J_2$, as shown in Fig.~\ref{fig:jscompare}.  A minimal network with kernel size 2 and $D_{\rm hidden}=16$ already tracks the qualitative deterioration of the solvable-point sVBS state, and systematically enlarging the receptive field and hidden dimension allows the ansatz to capture the extra entanglement. In this sense, the architecture remains perturbatively connected to the exactly solvable limit: the same local inductive bias that is exact at $J_1=J_2$ still organizes the representation efficiently away from that point.

\begin{figure}[t!]
\begin{center}
\includegraphics[width=0.99\linewidth]{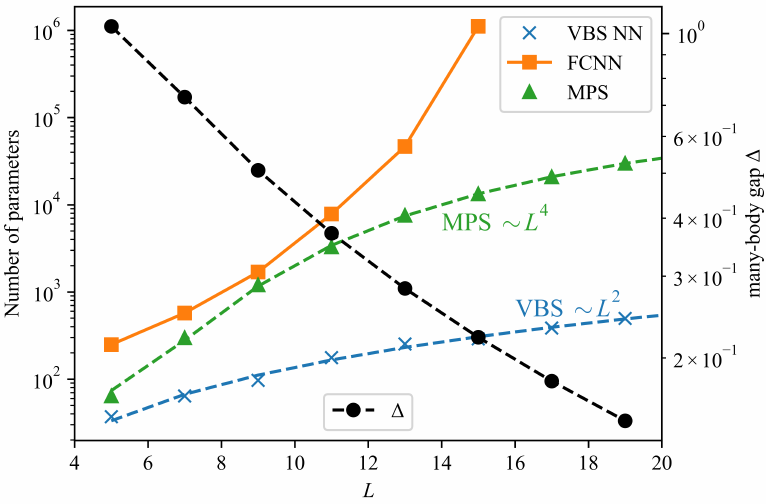}
\caption{\label{fig:scaling}
    The number of model parameters required by different ansatz states in the non-exact case $J_1=1$, $J_2=0.9$, $t_1=1$, and $t_2=0.5$. For the neural-network ansatz, the number of parameters is determined by requiring either the fidelity error or the relative energy error to be below $1\%$, averaged over $10$ runs. For MPS, the bond dimension is determined by an SVD cutoff of $10^{-12}$. The dashed curves are polynomial fits, giving approximately $L^4$ scaling for MPS and $L^2$ scaling for the VBS neural-network ansatz. The black dashed line indicates the many-body gap. 
}
\end{center}
\end{figure}
Figure~\ref{fig:scaling} further illustrates this distinction. Away from the exact point, the required bond dimension of the MPS grows rapidly, reflecting the increased entanglement of the ground state. By contrast, the neural-network ansatz reaches comparable accuracy with a more favorable parameter scaling. This suggests that the VBS-based architecture preserves the physically relevant structure even when exact solvability is lost.


Under supervised training against exact-diagonalization (ED) targets, the NQS maintains high fidelity with the ED ground state in this non-exact regime for system sizes ($L\le 19$) that are accessible to ED. Away from the solvable point, the architecture remains systematically improvable by increasing kernel size and hidden dimension, indicating that the inductive bias remains effective beyond exact solvability.

\section{Energy-Based Optimization via Variational Monte Carlo}\label{sec:5}

In Secs.~\ref{sec:2}--\ref{sec:4}, we focused on supervised compression of ED ground states into an NQS ansatz. 
For larger systems ($L\ge 21$), however, exact diagonalization is no longer feasible because of the exponential growth of the Hilbert space. In that regime, the natural alternative is variational Monte Carlo (VMC), in which the NQS is optimized directly by minimizing the variational energy rather than a supervised fidelity-based loss.

In VMC, one introduces a parametrized trial wavefunction $\ket{\Psi_\theta}$, represented here by a neural network with parameters $\theta$. For each many-body configuration $s$, the network outputs a complex amplitude $\Psi_\theta(s)$. The objective is to minimize the variational energy
\begin{equation}
    E_\theta
    =\frac{\langle \Psi_\theta|H|\Psi_\theta\rangle}
    {\langle\Psi_\theta|\Psi_\theta\rangle}
    =\sum_s P_\theta(s) E_{\mathrm{loc}}(s),
\end{equation}
where $P_\theta(s)=\frac{|\Psi_\theta(s)|^2}{\sum_s |\Psi_\theta(s)|^2}$
is the probability distribution induced by the trial wavefunction, and $E_{\mathrm{loc}}(s)=\sum_{s'} H_{s,s'}\frac{\Psi_\theta(s')}{\Psi_\theta(s)}$
is the local energy estimator. By sampling configurations $s$ from $P_\theta(s)$ using the Metropolis-Hastings algorithm \cite{Metropolis1953,Hastings1970}, the energy can be estimated as
\begin{equation}
    E_\theta \approx \frac{1}{N_s}\sum_{i=1}^{N_s} E_{\mathrm{loc}}(s_i), \qquad s_i \sim P_\theta(s).
\end{equation}

The energy gradient with respect to the wavefunction parameters is
\begin{equation}
    \begin{split}
        \frac{\partial E_\theta}{\partial \theta_k}
        &=2\left\langle
        E_{\mathrm{loc}}
        \frac{\partial_{\theta_k}\Psi_\theta(s)}{\Psi_\theta(s)}
        \right\rangle
        -2\langle E_{\mathrm{loc}}\rangle
        \left\langle
        \frac{\partial_{\theta_k}\Psi_\theta(s)}{\Psi_\theta(s)}
        \right\rangle,
    \end{split}
\end{equation}
where in the last step we used $\Psi_\theta(s)=\Psi_\theta^*(s)$ for the real-valued wavefunctions considered here.

A direct gradient update, $\delta\theta\propto -\nabla E$, treats parameter space as Euclidean. For quantum states, however, the natural notion of distance is the distance between wavefunctions themselves. This is encoded by the Fubini-Study metric \cite{Stokes2020quantumnatural,Provost1980Riemannian}. Writing
\begin{equation}
    \ket{\Psi(\theta+\delta\theta)}
    \approx
    \ket{\Psi(\theta)}
    +\sum_k \delta\theta_k
    \frac{\partial}{\partial\theta_k}\ket{\Psi(\theta)},
\end{equation}
one finds that the infinitesimal wavefunction distance takes the form $ds^2=\sum_{ij} S_{ij}\, d\theta_i d\theta_j,$
with metric tensor
\begin{equation}
    \begin{split}
        S_{ij}
        &= \frac{\langle\partial_i \Psi|\partial_j \Psi\rangle}
        {\langle\Psi|\Psi\rangle}
        -\frac{\langle\Psi|\partial_i \Psi\rangle
        \langle\partial_j \Psi|\Psi\rangle}
        {|\langle\Psi|\Psi\rangle|^2}.
    \end{split}
\end{equation}
This metric defines the natural geometry of the variational manifold. Instead of ordinary gradient descent, a natural update based on this geometry is the natural-gradient method, often implemented in VMC as stochastic reconfiguration (SR) \cite{PhysRevLett.80.4558,Stokes2020quantumnatural}, by solving $S\,\delta\boldsymbol{\theta}=\nabla E,$ followed by $\boldsymbol{\theta}\to \boldsymbol{\theta}-\eta\,\delta\boldsymbol{\theta}.$ In practice, the covariance matrix is estimated from Monte Carlo samples and regularized by a diagonal shift. The stochastic-reconfiguration estimator can be written in terms of the centered log-derivative matrix
\begin{equation}
    O_{ip} = \partial_p \ln \Psi_\theta(s_i) - \frac{1}{N_s}\sum_{j=1}^{N_s}\partial_p \ln \Psi_\theta(s_j),
\end{equation}
and the centered local-energy vector 
\begin{equation}
    [\vec{E}_{\mathrm{loc}}]_i = E_{\mathrm{loc}}(s_i) - \frac{1}{N_s}\sum_{j=1}^{N_s} E_{\mathrm{loc}}(s_j).
\end{equation}
The stochastic-reconfiguration step can then be written as
\begin{equation}
    2\lambda(S+\tau \mathbb{I}_{N_p})\delta\boldsymbol{\theta}
    =\nabla E
    \approx 2 O^T \vec{E}_{\mathrm{loc}},
\end{equation}
where $S=O^T O$ is the covariance matrix of the log-derivatives, $N_p$ is the number of variational parameters, $N_s$ is the number of Monte Carlo samples, $\tau$ is a regularization parameter, and $\lambda$ is a hyperparameter that sets the overall update scale.

The computational cost of VMC is controlled primarily by the number of parameters and the sample size. If $N_p$ denotes the number of model parameters and $N_s$ the number of samples used in one update, forming the covariance matrix scales as $\mathcal{O}(N_s N_p^2)$, while solving the linear system associated with stochastic reconfiguration scales as $\mathcal{O}(N_p^3)$. These costs quickly become substantial as the ansatz grows. In practice, matrix-free iterative solvers such as conjugate gradient give much longer VMC runtimes despite their better asymptotic scaling, and thus direct linear solvers are preferred for moderate parameter counts. 

In our implementation, the gradient is always obtained from the SR step in VMC, while the parameter update itself is performed in PyTorch \cite{paszke2019pytorch} using stochastic gradient descent with an adaptive learning rate; for comparison, Appendix~\ref{app:vmc-bench} reports scaling results without SR. We find that momentum-based optimizers such as Adam \cite{kingma2015adam} are generally less effective in this setting, because the momentum term can amplify the autocorrelation already present in successive Monte Carlo samples. In practice, the optimization may still become trapped in local minima, especially for unfavorable initializations. Running several independent seeds is therefore useful for identifying the best variational state. In the exactly solvable regime, reinitialization can also be necessary at large system size to avoid getting trapped near configurations with vanishing amplitude.

The distinction between supervised learning and VMC is therefore important. Supervised optimization is useful for benchmarking and for compressing ED ground states when exact targets are available, but it is restricted to small systems by the exponential growth of the Hilbert space. VMC provides the scalable route for training NQS directly at larger system sizes, since it optimizes the variational energy from samples rather than full enumeration. In practice, its accuracy is determined not only by the expressivity of the neural-network ansatz but also by the efficiency of the sampling and optimization procedure, including Monte Carlo autocorrelation and the stability of the SR updates \cite{Merali2026ParallelScanRNN}. In this sense, compact wavefunction representations and robust optimization are complementary ingredients, much as the effectiveness of DMRG relies on both the MPS ansatz and the sweep algorithm.

\begin{figure}[t!]
\begin{center}
\setlabel{pos=nw,fontsize=\large,labelbox=false}
\xincludegraphics[scale=0.7,label=a]{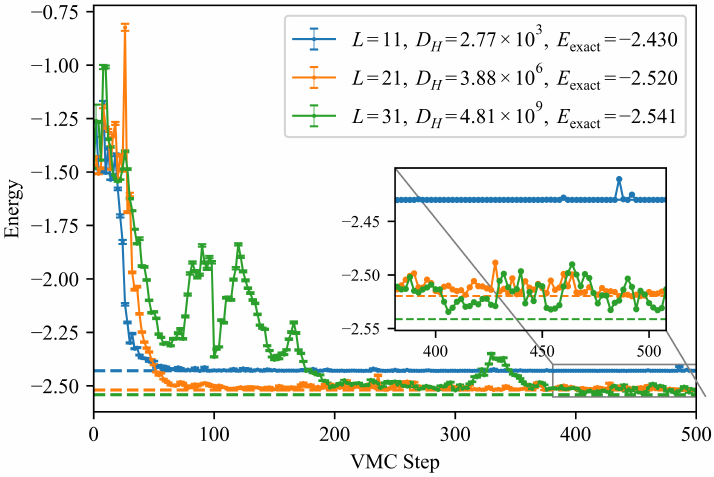}
\xincludegraphics[scale=0.68,label=b]{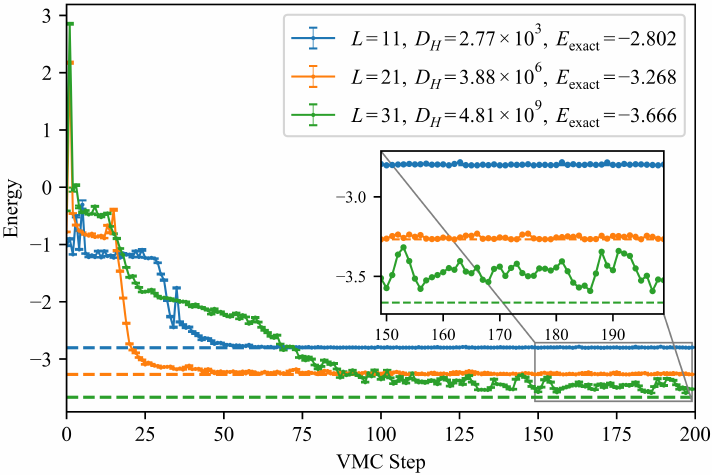}
\caption{\label{fig:vmc}
VMC results for the sVBS ground states. (a) sVBS ground states at the exactly solvable point, $t_2=0.5t_1$ and $J_1=J_2=0$. (b) Non-exact sVBS ground states at $t_2=0.5t_1$, $J_1=0.1$, and $J_2=0.09$. The energies are estimated using $64\times128=8192$ Monte Carlo samples. Dashed horizontal lines represent the benchmark value from DMRG. The learning rate is $0.05$ and regularization $\tau=0.1$.
}
\end{center}
\end{figure}

Fig. \ref{fig:vmc} shows the VMC results for both the exactly solvable and non-exact regimes. In both cases, the variational energy attains a value close to the benchmark result for the sample sizes and training times used here. As the system size increases, however, the energy error also increases. In the exactly solvable regime, the dominant difficulty originates from the hole sector represented by the fully connected part of the network. Even if the spin branch is fixed to reproduce the exact VBS wavefunction structure, learning the hole wavefunction becomes progressively harder as the system grows. This is consistent with the effective tight-binding description: for open chains with an odd number of sites, the finite-size gap of the hole sector decreases asymptotically as $1/L$, making optimization increasingly sensitive to sampling noise.

Interestingly, in the non-exact regime, improved optimization behavior is observed in terms of the number of VMC steps needed to reach the benchmark energy (determined via DMRG), even though the non-exact ground states are more entangled. At the same time, the increased entanglement demands a more expressive ansatz and therefore more parameters to reach high fidelity. The exactly solvable sVBS manifold provides a constrained Hilbert space that facilitates sampling, whereas away from the exactly solvable point the broader support of the wavefunction makes accurate VMC optimization increasingly expensive.

Overall, VMC remains viable in both exact and non-exact regimes, but its cost grows rapidly with system size. In the non-exact regime, the gapless nature \cite{Vidal2003Entanglement,Calabrese2004Entanglement,PhysRevB.111.035119} of the ground state and the associated logarithmic entanglement growth increase estimator variance and sensitivity to autocorrelation, so stable SR updates require larger sample sets and careful optimization. Consequently, the deterioration of VMC performance at larger $L$ should not be interpreted as a failure of the NQS architecture itself, but as the expected increase in sampling and optimization cost in the gapless regime.

\section{Translational Invariance and System-Size Extrapolation}\label{sec:6}

\begin{figure}[t!]
\begin{center}
\includegraphics[width=0.99\linewidth]{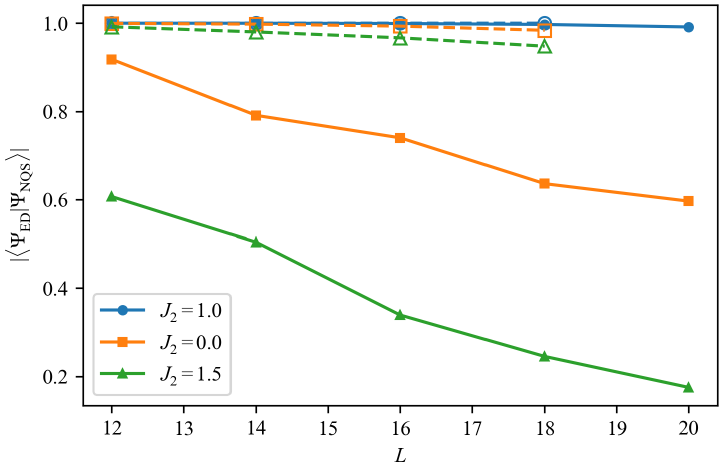}
\caption{\label{fig:extrapolate}
Extrapolation of neural networks trained on small system sizes, $L=2 \to 10$, to larger systems $L=12\to 20$ for several values of $J_2$ at fixed $J_1=1$. Solid curves show the fidelity of the extrapolated neural-network wavefunction on the larger systems, while dashed curves show the fidelity obtained by direct training at those system sizes. For the exactly solvable VBS point $J_1=J_2=1$, the network uses kernel size $2$ and hidden dimension $16$. For the gapless cases $J_1\neq J_2$, the network uses kernel size $5$ and hidden dimension $64$.
}
\end{center}
\end{figure}

Because the VBS-inspired NQS architecture successfully captures both the exactly solvable and non-exact sVBS regimes, it is natural to ask to what extent its spin sector retains a translationally organized structure beyond the exactly solvable point. In the pure VBS limit, the local singlet rule is translationally uniform in the bulk, and this is precisely the structure encoded by the convolutional part of the network. The question is whether the same locality and translational organization remain useful when the ground state is no longer exactly solvable.

To address this, we isolate the spin sector of the NQS and disregard the hole degree of freedom. The resulting problem is described by a quasi-one-dimensional spin Hamiltonian with an even number of sites and open boundaries,
\begin{equation}
\begin{split}
    H &= J_1 \sum_{\langle i,j\rangle} \mathbf{S}_i\cdot \mathbf{S}_j
    +J_2 \sum_{\langle \langle i,j\rangle\rangle,\; i,j\, \mathrm{odd}}
    \mathbf{S}_i\cdot \mathbf{S}_j,
\end{split}
\end{equation}
where $J_1,J_2\ge 0$ are the nearest-neighbor and next-nearest-neighbor antiferromagnetic couplings, respectively. At $J_1=J_2$, the VBS state is the exact ground state. When $J_1\neq J_2$, the system moves away from the exact point and develops increasing entanglement, becoming gapless in the thermodynamic limit.

The central observation is that the local singlet-pairing rule remains system-size independent. As a result, the NQS can be trained on small systems and then evaluated on larger ones without retraining, as shown in Fig.~\ref{fig:extrapolate}. At the exactly solvable VBS point, this extrapolation remains highly accurate: the infidelity increases only from $10^{-5}$ to $10^{-3}$ as the system size grows from $L=12$ to $L=20$. 
This behavior reflects the fact that the network has learned a truly local and size-independent rule, rather than merely interpolating within a fixed Hilbert space.

Away from the exactly solvable point, the situation changes qualitatively. Although the same architecture still captures the dominant local structure, the extrapolated fidelity decreases with system size. Direct training on the larger systems yields significantly better fidelity, indicating that the ground state is no longer fully determined by a translationally uniform local rule. In this sense, the degradation of extrapolation provides an indirect signature of the breakdown of the exact VBS structure and the emergence of additional long-range correlations.

This contrast highlights an important distinction between exact and non-exact regimes. In the exact case, the convolutional architecture together with geometric pooling is sufficient to propagate the learned local rule across system sizes while preserving the proper normalization structure. In the non-exact case, the same inductive bias remains useful, but the wavefunction requires additional flexibility to accommodate excess entanglement and the loss of strict translational organization.

The extrapolation behavior therefore provides a complementary diagnostic of the ansatz. It shows not only whether the NQS can fit a given system size, but also whether it has captured the underlying size-independent structure of the state. From this perspective, extrapolation may be viewed as a form of transfer learning: knowledge acquired on small systems is transferred to larger ones whenever the wavefunction is governed by stable local rules. The success of this transfer in the exactly solvable VBS regime, and its deterioration away from that regime, further supports the physical interpretation of the proposed architecture.

\section{Conclusion}\label{sec:7}

We have introduced a minimal neural-network quantum-state architecture for quasi-one-dimensional VBS and soliton VBS (sVBS) states. At the exactly solvable point of the quasi-one-dimensional $t$-$J$-like model, the proposed architecture can represent the sVBS ground state with near-unit fidelity while using a very small number of parameters. Away from the exactly solvable point, the same architecture remains systematically improvable: by increasing the kernel size and hidden dimension, it continues to capture the additional entanglement of the non-exact ground states with favorable parameter scaling.

Beyond the specific model studied here, our construction highlights several general design principles for neural-network quantum states. First, in the present VBS setting, translational structure is encoded primarily through the stride of the convolution rather than through a large kernel \cite{ybgv-35jm}. In particular, stride $2$ is sufficient to resolve the dimer pattern, while the kernel size mainly controls how much additional entanglement can be incorporated away from the exactly solvable limit. Second, geometric pooling provides a natural way to represent multiplicative wavefunction structure and the associated normalization across system sizes. Third, the choice of activation is important, with $\tanh(x^{2k+1})$ with $k>0$ found to be particularly effective for resolving the three-valued local rule $\{+1,0,-1\}$ that underlies the VBS wavefunction. Although such activations have vanishing derivatives at the origin, they are well-behaved in practice and improve the separation of the zero-amplitude sector from the nonzero sectors. Finally, when both spin and hole degrees of freedom are present, separating them within the architecture is essential: without this separation, generic brute-force architectures average over incompatible local structures and fail to reproduce the correct state.

These observations suggest that physically informed NQS architectures can provide a useful alternative to generic black-box networks. Although our construction is tailored to quasi-one-dimensional VBS-like states, the underlying principles may extend more broadly to systems with resonating valence-bond physics, constrained Hilbert spaces, or mixed local degrees of freedom. In particular, the spin-hole sector separation that encodes the fermion sign structure here applies directly to NQS for the broader $t$-$J$ and Hubbard families, where the entanglement between the spin background and mobile holes underlies the sign problem. In such settings, incorporating known local structure directly into the network architecture may lead to compact ansatz states that remain expressive beyond exactly solvable limits.

At the same time, our results show that representational efficiency alone is not sufficient for high-performance NQS calculations. As in tensor-network methods, overall success is determined jointly by the variational ansatz and the optimization algorithm: DMRG combines MPS with highly effective sweep updates, while NQS performance depends on the quality of VMC sampling and natural-gradient optimization. In the present problem, the proposed architecture provides a compact and accurate representation of the target states, and further gains will likely come from improved large-scale optimization, lower-variance sampling strategies, and more efficient implementations of SR.

This suggests a clear direction for future work. If more scalable optimization strategies can be developed, then physically structured neural-network ansatz states may become genuinely competitive for quantum many-body problems beyond the reach of conventional tensor-network methods. In this respect, it may be fruitful to further explore connections between VMC and reinforcement-learning-style optimization, especially in regimes where sampling efficiency is the dominant bottleneck.

\begin{acknowledgments}
The work at the UT Knoxville was primarily supported by the National Science Foundation Materials Research Science and Engineering Center program through the UT Knoxville Center for Advanced Materials and Manufacturing (DMR-2309083). The computation for this work was performed on the University of Tennessee Infrastructure for Scientific Applications and Advanced Computing (ISAAC) computational resources. We acknowledge the NetKet codebase as a learning  and inspiration resource for the codes~\cite{netket3:2022,netket2:2019}.
\end{acknowledgments}

\appendix

\begin{table*}[t!]
    \centering
    \renewcommand{\arraystretch}{1.5}
    \setlength\tabcolsep{8pt}
    \begin{tabular}{@{}lcccccc@{}}
        \toprule
        $\vb*{L}$ & 21 & 23 & 25 & 27 & 29 & 31 \\
        \midrule
        $\vb*{D_H}$ & $3{,}879{,}876$ & $1.62\times 10^7$ & $6.76\times 10^7$ & $2.81\times 10^8$ & $1.16\times 10^9$ & $4.81\times 10^9$ \\
        $E_{\mathrm{exact}}$ & $-2.519684$ & $-2.526143$ & $-2.531216$ & $-2.535273$ & $-2.538568$ & $-2.541281$ \\
        $E_{\mathrm{vmc}}$ & $-2.514891$ & $-2.523682$ & $-2.524803$ & $-2.5232678$ & $-2.521457$ & $-2.535425$ \\
        \bottomrule
    \end{tabular}
    \caption{
    Fixed-budget VMC benchmark at $t_1=1$ and $t_2=0.5$ with $N_s=10{,}000$ samples and $100$ optimization epochs. The spin branch is pretrained and frozen, so VMC optimizes only the hole sector.
    }
    \label{tab:vmc_pretrain}
\end{table*}

\begin{table*}[t!]
    \centering
    \renewcommand{\arraystretch}{1.5}
    \setlength\tabcolsep{8pt}
    \begin{tabular}{@{}lcccccc@{}}
        \toprule
        Setting & 21 & 23 & 25 & 27 & 29 & 31 \\
        \midrule
        $E_{\mathrm{exact}}$ & $-2.519684$ & $-2.526143$ & $-2.531216$ & $-2.535273$ & $-2.538568$ & $-2.541281$ \\
        \midrule
        $E_{\mathrm{vmc}}$ (epoch $1000$, $N_s=10^4$) & $-2.520858$ & $-2.425977$ & $-2.254782$ & $-2.280139$ & $-2.175475$ & $-2.257368$ \\
        $E_{\mathrm{vmc}}$ (epoch $2000$, $N_s=10^4$) & $-2.507296$ & $-2.244623$ & $-2.163037$ & $-2.267434$ & $-2.176712$ & $-2.218174$ \\
        $E_{\mathrm{vmc}}$ (epoch $3000$, $N_s=10^4$) & $-2.498387$ & $-2.406371$ & $-2.239218$ & $-2.269608$ & $-2.183908$ & $-2.206605$ \\
        $E_{\mathrm{vmc}}$ (epoch $100$, $N_s=10^5$) & $-2.499968$ & $-2.505271$ & $-2.246792$ & $-2.256597$ & $-2.182826$ & $-2.193448$ \\
        $E_{\mathrm{vmc}}$ (hole sampler, $N_s=10^4$) & $-2.510174$ & $-2.468095$ & $-2.289946$ & $-2.296831$ & $-2.186189$ & $-2.215386$ \\
        \bottomrule
    \end{tabular}
    \caption{
    Large-system VMC sensitivity study at $t_1=1$ and $t_2=0.5$ with hidden dimension $64$, using direct gradient optimization without SR. Rows compare optimization budgets and sampling strategies (random-two-site updates versus nearest-neighbor hole moves). For reference, an additional benchmark is $E_{\mathrm{exact}}(L=101)=-2.55856$.
    }
    \label{tab:app_vmc_scaling}
\end{table*}

\section{Hopping Matrices on a Single $\Delta$ Triangle}\label{app:hopping}

On a single $\Delta$ triangle, the hopping term $h_\alpha^t$ acts within the sector containing one hole and two spins. Decomposing the two-spin sector into singlet ($S=0$) and triplet ($S=1$) channels, $h_\alpha^t$ reduces to a $3\times 3$ matrix in each channel, with rows and columns labeled by the position of the hole on the three sites of the triangle,
\begin{equation}
    \begin{split}
        h_\alpha^t(\mathbf{S}=0) &=
        \begin{pmatrix}
            0 & -t_1 & -t_2 \\
            -t_1 & 0 & -t_1 \\
            -t_2 & -t_1 & 0
        \end{pmatrix},\\
        h_\alpha^t(\mathbf{S}=1) &=
        \begin{pmatrix}
            0 & -t_1 & t_2 \\
            -t_1 & 0 & -t_1 \\
            t_2 & -t_1 & 0
        \end{pmatrix}.
    \end{split}
\end{equation}
The sign structure of the off-diagonal matrix elements follows from fermionic statistics. The lowest eigenvalues are
\begin{equation}
    \lambda_{\min}(S=0)=\frac{-t_2-\sqrt{8t_1^2+t_2^2}}{2},
\end{equation}
and
\begin{equation}
    \lambda_{\min}(S=1)=
    \min\left[
        -t_2,\,
        \frac{t_2-\sqrt{8t_1^2+t_2^2}}{2}
    \right].
\end{equation}
For $t_1,t_2>0$, one has $\lambda_{\min}(S=0)<\lambda_{\min}(S=1)$, so the kinetic term favors the singlet sector on each triangle.

\section{Additional VMC Convergence and Scaling Data}\label{app:vmc-bench}

This appendix summarizes additional VMC benchmarks used to characterize convergence and large-system scaling beyond the main figures.  
Table~\ref{tab:vmc_pretrain} reports fixed-budget runs with $N_s=10^4$ samples and $100$ optimization epochs. In these runs, the neural network is first pretrained to learn the local spin rule, and the spin branch is then frozen during VMC so that optimization targets only the hole wavefunction. Even under this favorable setup, the energy error grows with system size, consistent with increasing sampling noise and optimization difficulty in the hole sector at larger $L$.

Table~\ref{tab:app_vmc_scaling} compares optimization settings (epoch budget, sample size, and sampler choice) for the same Hamiltonian parameters. In these tests, stochastic reconfiguration (SR) is not used; instead, the raw energy gradient is optimized with Adam. Without SR preconditioning, optimization is more sensitive to Monte Carlo noise, and the error increases substantially with system size. The final row compares a random-two-site update sampler with a nearest-neighbor hole-moving sampler. The hole-moving sampler explores the relevant configuration manifold more efficiently in the exactly solvable regime and yields improved large-$L$ energies, whereas the random-two-site sampler is generally more robust away from the exactly solvable point.

Overall, these results show that VMC cost rises rapidly with system size, and while tuning optimization and sampling improves performance, it does not fully eliminate the large-$L$ bottleneck.

\section*{Code Availability}
The code used to generate the numerical results in this work is publicly available at
\href{https://github.com/angkun-research/NQSpy}{https://github.com/angkun-research/NQSpy}.

\bibliography{reference.bib}

@article{glittum2025resonant,
  title={A resonant valence bond spin liquid in the dilute limit of doped frustrated Mott insulators},
  author={Glittum, Cecilie and {\v{S}}trkalj, Antonio and Prabhakaran, Dharmalingam and Goddard, Paul A and Batista, Cristian D and Castelnovo, Claudio},
  journal={Nat. Phys.},
  pages={1--6},
  year={2025},
  publisher={Nature Publishing Group UK London},
 url={https://www.nature.com/articles/s41567-025-02923-8}
}

@article{PhysRevB.107.L140401,
  title = {Exact hole-induced resonating-valence-bond ground state in certain $U=\ensuremath{\infty}$ Hubbard models},
  author = {Kim, Kyung-Su},
  journal = {Phys. Rev. B},
  volume = {107},
  issue = {14},
  pages = {L140401},
  numpages = {5},
  year = {2023},
  month = {Apr},
  publisher = {American Physical Society},
  doi = {10.1103/PhysRevB.107.L140401},
  url = {https://link.aps.org/doi/10.1103/PhysRevB.107.L140401}
}

@article{White1992DMRG,
  title = {Density matrix formulation for quantum renormalization groups},
  author = {White, Steven R.},
  journal = {Phys. Rev. Lett.},
  volume = {69},
  issue = {19},
  pages = {2863--2866},
  year = {1992},
  doi = {10.1103/PhysRevLett.69.2863},
  url = {https://link.aps.org/doi/10.1103/PhysRevLett.69.2863}
}

@article{Schollwock2011MPS,
  title = {The density-matrix renormalization group in the age of matrix product states},
  author = {Schollwock, Ulrich},
  journal = {Ann. Phys.},
  volume = {326},
  number = {1},
  pages = {96--192},
  year = {2011},
  doi = {10.1016/j.aop.2010.09.012},
  url = {https://doi.org/10.1016/j.aop.2010.09.012}
}

@article{Carleo2017NQS,
  title = {Solving the quantum many-body problem with artificial neural networks},
  author = {Carleo, Giuseppe and Troyer, Matthias},
  journal = {Science},
  volume = {355},
  number = {6325},
  pages = {602--606},
  year = {2017},
  doi = {10.1126/science.aag2302},
  url = {https://doi.org/10.1126/science.aag2302}
}

@article{Affleck1987AKLT,
  title = {Rigorous results on valence-bond ground states in antiferromagnets},
  author = {Affleck, Ian and Kennedy, Tom and Lieb, Elliott H. and Tasaki, Hal},
  journal = {Phys. Rev. Lett.},
  volume = {59},
  issue = {7},
  pages = {799--802},
  year = {1987},
  doi = {10.1103/PhysRevLett.59.799},
  url = {https://doi.org/10.1103/PhysRevLett.59.799}
}

@article{Anderson1973RVB,
  title = {Resonating valence bonds: A new kind of insulator?},
  author = {Anderson, P. W.},
  journal = {Mater. Res. Bull.},
  volume = {8},
  number = {2},
  pages = {153--160},
  year = {1973},
  doi = {10.1016/0025-5408(73)90167-0},
  url = {https://doi.org/10.1016/0025-5408(73)90167-0}
}

@article{Metropolis1953,
  title = {Equation of State Calculations by Fast Computing Machines},
  author = {Metropolis, Nicholas and Rosenbluth, Arianna W. and Rosenbluth, Marshall N. and Teller, Augusta H. and Teller, Edward},
  journal = {J. Chem. Phys.},
  volume = {21},
  number = {6},
  pages = {1087--1092},
  year = {1953},
  doi = {10.1063/1.1699114},
  url = {https://doi.org/10.1063/1.1699114}
}

@article{Hastings1970,
  title = {Monte Carlo sampling methods using Markov chains and their applications},
  author = {Hastings, W. K.},
  journal = {Biometrika},
  volume = {57},
  number = {1},
  pages = {97--109},
  year = {1970},
  doi = {10.1093/biomet/57.1.97},
  url = {https://doi.org/10.1093/biomet/57.1.97}
}

@article{Vidal2003Entanglement,
  title = {Entanglement in quantum critical phenomena},
  author = {Vidal, Guifre and Latorre, Jose I. and Rico, Enrique and Kitaev, Alexei},
  journal = {Phys. Rev. Lett.},
  volume = {90},
  issue = {22},
  pages = {227902},
  year = {2003},
  doi = {10.1103/PhysRevLett.90.227902},
  url = {https://doi.org/10.1103/PhysRevLett.90.227902}
}

@article{Calabrese2004Entanglement,
  title = {Entanglement entropy and quantum field theory},
  author = {Calabrese, Pasquale and Cardy, John},
  journal = {J. Stat. Mech.: Theory Exp.},
  volume = {2004},
  number = {06},
  pages = {P06002},
  year = {2004},
  doi = {10.1088/1742-5468/2004/06/P06002},
  url = {https://doi.org/10.1088/1742-5468/2004/06/P06002}
}

@article{Dosovitskiy2021ViT,
  title = {An Image is Worth 16x16 Words: Transformers for Image Recognition at Scale},
  author = {Dosovitskiy, Alexey and Beyer, Lucas and Kolesnikov, Alexander and Weissenborn, Dirk and Zhai, Xiaohua and Unterthiner, Thomas and Dehghani, Mostafa and Minderer, Matthias and Heigold, Georg and Gelly, Sylvain and Uszkoreit, Jakob and Houlsby, Neil},
  journal = {arXiv:2010.11929},
  year = {2021},
  url = {https://arxiv.org/abs/2010.11929}
}

@article{PhysRevLett.80.4558,
  title = {Green Function Monte Carlo with Stochastic Reconfiguration},
  author = {Sorella, Sandro},
  journal = {Phys. Rev. Lett.},
  volume = {80},
  issue = {20},
  pages = {4558--4561},
  numpages = {0},
  year = {1998},
  month = {May},
  publisher = {American Physical Society},
  doi = {10.1103/PhysRevLett.80.4558},
  url = {https://link.aps.org/doi/10.1103/PhysRevLett.80.4558}
}

@article{Stokes2020quantumnatural,
  doi = {10.22331/q-2020-05-25-269},
  url = {https://doi.org/10.22331/q-2020-05-25-269},
  title = {Quantum {N}atural {G}radient},
  author = {Stokes, James and Izaac, Josh and Killoran, Nathan and Carleo, Giuseppe},
  journal = {{Quantum}},
  issn = {2521-327X},
  publisher = {{Verein zur F{\"{o}}rderung des Open Access Publizierens in den Quantenwissenschaften}},
  volume = {4},
  pages = {269},
  month = may,
  year = {2020}
}

@article{Provost1980Riemannian,
  author = {Provost, J. P. and Vallee, G.},
  title = {Riemannian structure on manifolds of quantum states},
  journal = {Commun. Math. Phys.},
  volume = {76},
  number = {3},
  pages = {289--301},
  year = {1980},
  doi = {10.1007/BF02193559},
  url = {https://doi.org/10.1007/BF02193559}
}

@article{PhysRevB.111.035119,
  title = {Disentangling interacting systems with fermionic Gaussian circuits: Application to quantum impurity models},
  author = {Wu, Ang-Kun and Kloss, Benedikt and Krinitsin, Wladislaw and Fishman, Matthew T. and Pixley, J. H. and Stoudenmire, E. M.},
  journal = {Phys. Rev. B},
  volume = {111},
  issue = {3},
  pages = {035119},
  numpages = {18},
  year = {2025},
  month = {Jan},
  publisher = {American Physical Society},
  doi = {10.1103/PhysRevB.111.035119},
  url = {https://link.aps.org/doi/10.1103/PhysRevB.111.035119}
}

@article{wu2025modeling,
  title = {Modeling quantum geometry for fractional Chern insulators with unsupervised learning},
  author = {Wu, Ang-Kun and Primeau, Louis and Zhang, Jingtao and Sun, Kai and Zhang, Yang and Lin, Shi-Zeng},
  journal = {npj Comput. Mater.},
  year = {2026},
  month = {May},
  doi = {10.1038/s41524-026-02155-1},
  url = {https://www.nature.com/articles/s41524-026-02155-1}
}

@article{StoudenmireWhite2012,
  title = {Studying Two-Dimensional Systems with the Density Matrix Renormalization Group},
  author = {Stoudenmire, E. Miles and White, Steven R.},
  journal = {Annu. Rev. Condens. Matter Phys.},
  volume = {3},
  pages = {111--128},
  year = {2012},
  doi = {10.1146/annurev-conmatphys-020911-125018},
  url = {https://doi.org/10.1146/annurev-conmatphys-020911-125018}
}

@article{PerezGarcia2007MPS,
  title = {Matrix Product State Representations},
  author = {Perez-Garcia, David and Verstraete, Frank and Wolf, Michael M. and Cirac, J. Ignacio},
  journal = {Quantum Inf. Comput.},
  volume = {7},
  number = {5-6},
  pages = {401--430},
  year = {2007},
  url = {https://arxiv.org/abs/quant-ph/0608197}
}

@article{Tagliacozzo2008FiniteEntanglement,
  title = {Scaling of entanglement support for matrix product states},
  author = {Tagliacozzo, Luca and de Oliveira, Thiago R. and Iblisdir, Sofyan and Latorre, Jose I.},
  journal = {Phys. Rev. B},
  volume = {78},
  pages = {024410},
  year = {2008},
  doi = {10.1103/PhysRevB.78.024410},
  url = {https://doi.org/10.1103/PhysRevB.78.024410}
}

@article{Pollmann2009FiniteEntanglement,
  title = {Theory of finite-entanglement scaling at one-dimensional quantum critical points},
  author = {Pollmann, Frank and Mukerjee, Subroto and Turner, Ari M. and Moore, Joel E.},
  journal = {Phys. Rev. Lett.},
  volume = {102},
  pages = {255701},
  year = {2009},
  doi = {10.1103/PhysRevLett.102.255701},
  url = {https://doi.org/10.1103/PhysRevLett.102.255701}
}

@article{CiracVerstraete2009,
  title = {Renormalization and tensor product states in spin chains and lattices},
  author = {Cirac, J. Ignacio and Verstraete, Frank},
  journal = {J. Phys. A: Math. Theor.},
  volume = {42},
  pages = {504004},
  year = {2009},
  doi = {10.1088/1751-8113/42/50/504004},
  url = {https://doi.org/10.1088/1751-8113/42/50/504004}
}

@article{Orus2014,
  title = {A practical introduction to tensor networks: Matrix product states and projected entangled pair states},
  author = {Or{\'u}s, Rom{\'a}n},
  journal = {J. Phys. A: Math. Theor.},
  volume = {47},
  pages = {424004},
  year = {2014},
  doi = {10.1088/1751-8113/47/42/424004},
  url = {https://doi.org/10.1088/1751-8113/47/42/424004}
}

@article{VerstraeteCirac2004PEPS,
  title = {Renormalization algorithms for Quantum-Many Body Systems in two and higher dimensions},
  author = {Verstraete, Frank and Cirac, J. Ignacio},
  journal = {arXiv:cond-mat/0407066},
  year = {2004},
  url = {https://arxiv.org/abs/cond-mat/0407066}
}

@article{Carrasquilla2017,
  title = {Machine learning phases of matter},
  author = {Carrasquilla, Juan and Melko, Roger G.},
  journal = {Nat. Phys.},
  volume = {13},
  pages = {431--434},
  year = {2017},
  doi = {10.1038/nphys4035},
  url = {https://doi.org/10.1038/nphys4035}
}

@article{Carleo2019ML,
  title = {Machine learning and the physical sciences},
  author = {Carleo, Giuseppe and Cirac, Ignacio and Cranmer, Kyle and Daudet, Laurent and Schuld, Maria and Tishby, Naftali and Vogt-Maranto, Leslie and Zdeborov{\'a}, Lenka},
  journal = {Rev. Mod. Phys.},
  volume = {91},
  pages = {045002},
  year = {2019},
  doi = {10.1103/RevModPhys.91.045002},
  url = {https://doi.org/10.1103/RevModPhys.91.045002}
}

@article{Mehta2019,
  title = {A high-bias, low-variance introduction to Machine Learning for physicists},
  author = {Mehta, Pankaj and Bukov, Marin and Wang, Ching-Hong and Day, Alexandre G. R. and Richardson, Clint and Fisher, Charles K. and Schwab, David J.},
  journal = {Phys. Rep.},
  volume = {810},
  pages = {1--124},
  year = {2019},
  doi = {10.1016/j.physrep.2019.03.001},
  url = {https://doi.org/10.1016/j.physrep.2019.03.001}
}

@article{GaoDuan2017,
  title = {Efficient representation of quantum many-body states with deep neural networks},
  author = {Gao, Xun and Duan, Lu-Ming},
  journal = {Nat. Commun.},
  volume = {8},
  pages = {662},
  year = {2017},
  doi = {10.1038/s41467-017-00705-2},
  url = {https://doi.org/10.1038/s41467-017-00705-2}
}

@article{Deng2017,
  title = {Quantum Entanglement in Neural Network States},
  author = {Deng, Dong-Ling and Li, Xiaopeng and Das Sarma, S.},
  journal = {Phys. Rev. X},
  volume = {7},
  pages = {021021},
  year = {2017},
  doi = {10.1103/PhysRevX.7.021021},
  url = {https://doi.org/10.1103/PhysRevX.7.021021}
}

@article{Nomura2017,
  title = {Restricted Boltzmann machine learning for solving strongly correlated quantum systems},
  author = {Nomura, Yusuke and Darmawan, Andrew S. and Yamaji, Youhei and Imada, Masatoshi},
  journal = {Phys. Rev. B},
  volume = {96},
  pages = {205152},
  year = {2017},
  doi = {10.1103/PhysRevB.96.205152},
  url = {https://doi.org/10.1103/PhysRevB.96.205152}
}

@article{Chen2018,
  title = {Equivalence of restricted Boltzmann machines and tensor network states},
  author = {Chen, Ji-Yao and Cheng, Song and Xie, Haidong and Wang, Lei and Xiang, Tao},
  journal = {Phys. Rev. B},
  volume = {97},
  pages = {085104},
  year = {2018},
  doi = {10.1103/PhysRevB.97.085104},
  url = {https://doi.org/10.1103/PhysRevB.97.085104}
}

@article{Glasser2018,
  title = {Neural-Network Quantum States, String-Bond States, and Chiral Topological States},
  author = {Glasser, Ivan and Pancotti, Nicola and August, Max and Rodr{\'i}guez, Ivan D. and Cirac, J. Ignacio},
  journal = {Phys. Rev. X},
  volume = {8},
  pages = {011006},
  year = {2018},
  doi = {10.1103/PhysRevX.8.011006},
  url = {https://doi.org/10.1103/PhysRevX.8.011006}
}

@article{LuoClark2019,
  title = {Backflow Transformations via Neural Networks for Quantum Many-Body Wave Functions},
  author = {Luo, Di and Clark, Bryan K.},
  journal = {Phys. Rev. Lett.},
  volume = {122},
  pages = {226401},
  year = {2019},
  doi = {10.1103/PhysRevLett.122.226401},
  url = {https://doi.org/10.1103/PhysRevLett.122.226401}
}

@article{Sharir2020,
  title = {Deep Autoregressive Models for the Efficient Variational Simulation of Many-Body Quantum Systems},
  author = {Sharir, Or and Ovadia, Ben and Neuberg, Gil and Falkovich, Ido and Kiss, Oren and Nevo, Netanel and Lavy, Yotam and Vardi, Jacob and Shkolnik, Tamar and Cohen, Mati and Geiger, Mario and Goldshlager, Naftali and Dagan, Yuval and Carleo, Giuseppe and Shashua, Amnon},
  journal = {Phys. Rev. Lett.},
  volume = {124},
  pages = {020503},
  year = {2020},
  doi = {10.1103/PhysRevLett.124.020503},
  url = {https://doi.org/10.1103/PhysRevLett.124.020503}
}

@article{HibatAllah2020,
  title = {Recurrent neural network wave functions},
  author = {Hibat-Allah, Mohamed and Ganahl, Martin and Melko, Roger and Carleo, Giuseppe and Choo, Kenny},
  journal = {Phys. Rev. Research},
  volume = {2},
  pages = {023358},
  year = {2020},
  doi = {10.1103/PhysRevResearch.2.023358},
  url = {https://doi.org/10.1103/PhysRevResearch.2.023358}
}

@article{Hermann2020,
  title = {Deep-neural-network solution of the electronic Schr{\"o}dinger equation},
  author = {Hermann, Jan and Sch{\"a}tzle, Zeno and No{\'e}, Frank},
  journal = {Nat. Chem.},
  volume = {12},
  pages = {891--897},
  year = {2020},
  doi = {10.1038/s41557-020-0544-y},
  url = {https://doi.org/10.1038/s41557-020-0544-y}
}

@article{Pfau2020,
  title = {Ab-Initio Solution of the Many-Electron Schr{\"o}dinger Equation with Deep Neural Networks},
  author = {Pfau, David and Spencer, James S. and Matthews, Alexander G. de G. and Foulkes, W. M. C.},
  journal = {Phys. Rev. Research},
  volume = {2},
  pages = {033429},
  year = {2020},
  doi = {10.1103/PhysRevResearch.2.033429},
  url = {https://doi.org/10.1103/PhysRevResearch.2.033429}
}

@article{ChenHeyl2024,
  title = {Empowering deep neural quantum states through efficient optimization},
  author = {Chen, Ao and Heyl, Markus},
  journal = {Nat. Phys.},
  volume = {20},
  pages = {1476--1481},
  year = {2024},
  doi = {10.1038/s41567-024-02566-1},
  url = {https://doi.org/10.1038/s41567-024-02566-1}
}

@article{Rende2024LargeScaleNQS,
  title = {A simple linear algebra identity to optimize Large-Scale Neural Network Quantum States},
  author = {Rende, Riccardo and Viteritti, Luciano Loris and Bardone, Lorenzo and Becca, Federico and Goldt, Sebastian},
  journal = {Commun. Phys.},
  volume = {7},
  pages = {260},
  year = {2024},
  doi = {10.1038/s42005-024-01732-4},
  url = {https://doi.org/10.1038/s42005-024-01732-4}
}

@article{Machaczek2025FractonNQS,
  title = {Neural Quantum State Study of Fracton Models},
  author = {Machaczek, Marc and Pollet, Lode and Liu, Ke},
  journal = {SciPost Phys.},
  volume = {18},
  number = {3},
  pages = {112},
  year = {2025},
  doi = {10.21468/SciPostPhys.18.3.112},
  url = {https://doi.org/10.21468/SciPostPhys.18.3.112}
}

@article{LuoHalverson2023InfiniteNQS,
  title = {Infinite Neural Network Quantum States: Entanglement and Training Dynamics},
  author = {Luo, Di and Halverson, James},
  journal = {Mach. Learn.: Sci. Technol.},
  volume = {4},
  pages = {025038},
  year = {2023},
  doi = {10.1088/2632-2153/ace02f},
  url = {https://doi.org/10.1088/2632-2153/ace02f}
}

@article{Inui2021DeterminantFree,
  title = {Determinant-free fermionic wave function using feed-forward neural networks},
  author = {Inui, Koji and Kato, Yasuyuki and Motome, Yukitoshi},
  journal = {Phys. Rev. Research},
  volume = {3},
  pages = {043126},
  year = {2021},
  doi = {10.1103/PhysRevResearch.3.043126},
  url = {https://doi.org/10.1103/PhysRevResearch.3.043126}
}

@article{Paul2026EntanglementBound,
  title = {Bound on Entanglement in Neural Quantum States},
  author = {Paul, Nisarga},
  journal = {Phys. Rev. Lett.},
  volume = {136},
  issue = {12},
  pages = {120403},
  numpages = {7},
  year = {2026},
  month = {Mar},
  publisher = {American Physical Society},
  doi = {10.1103/rpj5-cns6},
  url = {https://link.aps.org/doi/10.1103/rpj5-cns6}
}

@article{Cortes2026BasisDependence,
  title = {Basis dependence of Neural Quantum States for the Transverse Field Ising Model},
  author = {Cortes, Ronald Santiago and Shankar, Aravindh S. and Dalmonte, Marcello and Verdel, Roberto and Niggemann, Nils},
  journal = {SciPost Phys. Core},
  volume = {9},
  pages = {027},
  year = {2026},
  doi = {10.21468/SciPostPhysCore.9.2.027},
  url = {https://doi.org/10.21468/SciPostPhysCore.9.2.027}
}

@article{Roth2025HubbardSuperconductivity,
  title = {Superconductivity in the two-dimensional Hubbard model revealed by neural quantum states},
  author = {Roth, Christopher and Chen, Ao and Sengupta, Anirvan and Georges, Antoine},
  journal = {arXiv:2511.07566},
  year = {2025},
  url = {https://arxiv.org/abs/2511.07566}
}

@article{Yamazaki2026TransformerNQS,
  title = {Physics-inspired transformer quantum states via latent imaginary-time evolution},
  author = {Yamazaki, Kimihiro and Sakata, Itsushi and Konishi, Takuya and Kawahara, Yoshinobu},
  journal = {arXiv:2602.03031},
  year = {2026},
  url = {https://arxiv.org/abs/2602.03031}
}

@article{Hul2026GrandCanonicalNQS,
  title = {Neural network quantum states in the grand canonical ensemble},
  author = {Hul, Anton and Medvidovi{\'c}, Matija and Carrasquilla, Juan},
  journal = {arXiv:2605.07779},
  year = {2026},
  url = {https://arxiv.org/abs/2605.07779}
}

@article{Merali2026ParallelScanRNN,
  title = {Parallel Scan Recurrent Neural Quantum States for Scalable Variational Monte Carlo},
  author = {Merali, Ejaaz and Hibat-Allah, Mohamed and Kohandel, Mohammad and Scalettar, Richard T. and Khatami, Ehsan},
  journal = {arXiv:2605.13807},
  year = {2026},
  url = {https://arxiv.org/abs/2605.13807}
}

@article{Gu2026ParetoBackflow,
  title = {Pareto Frontier of Neural Quantum States: Scalable, Affordable, and Accurate Convolutional Backflow for Strongly Correlated Lattice Fermions},
  author = {Gu, Yuntian and Han, Zeyao and Li, Wenrui and Xiao, Zhiyu and Xiang, Tao and Qin, Mingpu and Wang, Liwei and Lv, Dingshun},
  journal = {arXiv:2604.25775},
  year = {2026},
  url = {https://arxiv.org/abs/2604.25775}
}

@incollection{paszke2019pytorch,
  title = {PyTorch: An Imperative Style, High-Performance Deep Learning Library},
  author = {Paszke, Adam and Gross, Sam and Massa, Francisco and Lerer, Adam and Bradbury, James and Chanan, Gregory and Killeen, Trevor and Lin, Zeming and Gimelshein, Natalia and Antiga, Luca and Desmaison, Alban and Kopf, Andreas and Yang, Edward and DeVito, Zachary and Raison, Martin and Tejani, Alykhan and Chilamkurthy, Sasank and Steiner, Benoit and Fang, Lu and Bai, Junjie and Chintala, Soumith},
  booktitle = {Advances in Neural Information Processing Systems 32},
  editor = {Wallach, H. and Larochelle, H. and Beygelzimer, A. and d'Alche-Buc, F. and Fox, E. and Garnett, R.},
  pages = {8024--8035},
  year = {2019},
  publisher = {Curran Associates, Inc.}
}

@article{barghathi:2026rc,
      title={{Detection of a R\'enyi Index Dependent Transition in Entanglement
             Entropy Scaling}}, 
      author={Hatem Barghathi and Adrian {Del Maestro}},
      year={2026},
      journal={arXiv:2512.24533},
      volume = {},
      pages = {},
      url={https://arxiv.org/abs/2512.24533}, 
      doi={10.48550/arXiv.2512.24533}
}

@article{Choo2018SymmetriesExcitedNQS,
  title = {Symmetries and many-body excited states with neural-network quantum states},
  author = {Choo, Kenny and Carleo, Giuseppe and Regnault, Nicolas and Neupert, Titus},
  journal = {Phys. Rev. Lett.},
  volume = {121},
  issue = {16},
  pages = {167204},
  year = {2018},
  doi = {10.1103/PhysRevLett.121.167204},
  url = {https://doi.org/10.1103/PhysRevLett.121.167204}
}

@article{Nomura2021RestoringSymmetry,
  title = {Helping restricted Boltzmann machines with quantum-state representation by restoring symmetry},
  author = {Nomura, Yusuke},
  journal = {J. Phys.: Condens. Matter},
  volume = {33},
  number = {17},
  pages = {174003},
  year = {2021},
  doi = {10.1088/1361-648X/abe268},
  url = {https://doi.org/10.1088/1361-648X/abe268}
}

@article{McClean2018BarrenPlateaus,
  title = {Barren plateaus in quantum neural network training landscapes},
  author = {McClean, Jarrod R. and Boixo, Sergio and Smelyanskiy, Vadim N. and Babbush, Ryan and Neven, Hartmut},
  journal = {Nat. Commun.},
  volume = {9},
  pages = {4812},
  year = {2018},
  doi = {10.1038/s41467-018-07090-4},
  url = {https://doi.org/10.1038/s41467-018-07090-4}
}

@article{Cerezo2021CostFunctionBP,
  title = {Cost function dependent barren plateaus in shallow parametrized quantum circuits},
  author = {Cerezo, M. and Sone, A. and Volkoff, T. and Cincio, L. and Coles, P. J.},
  journal = {Nat. Commun.},
  volume = {12},
  pages = {1791},
  year = {2021},
  doi = {10.1038/s41467-021-21728-w},
  url = {https://doi.org/10.1038/s41467-021-21728-w}
}

@article{Larocca2025BarrenPlateausReview,
  title = {Barren Plateaus in Variational Quantum Computing},
  author = {Larocca, Martin and Thanasilp, Supanut and Wang, Samson and Sharma, Kunal and Biamonte, Jacob and Coles, Patrick J. and Cincio, Lukasz and McClean, Jarrod R. and others},
  journal = {Nat. Rev. Phys.},
  volume = {7},
  pages = {174--189},
  year = {2025},
  doi = {10.1038/s42254-025-00813-9},
  url = {https://doi.org/10.1038/s42254-025-00813-9}
}

@Article{Rausch2023QuantumSpinSpiral,
	title={{Quantum spin spiral ground state of the ferrimagnetic sawtooth chain}},
	author={Roman Rausch and Matthias Peschke and Cassian Plorin and Jürgen Schnack and Christoph Karrasch},
	journal={SciPost Phys.},
	volume={14},
	pages={052},
	year={2023},
	publisher={SciPost},
	doi={10.21468/SciPostPhys.14.3.052},
	url={https://scipost.org/10.21468/SciPostPhys.14.3.052},
}

@article{Jiang2015Sawtooth,
author = {Jiang, Jian-Jun and Liu, Yong-Jun and Tang, Fei and Yang, Cui-Hong and Sheng, Yu-Bo},
title = {Analytical and numerical studies of the one-dimensional sawtooth chain},
journal = {Physica B: Condensed Matter},
volume = {463},
pages = {30--38},
year = {2015},
doi = {10.1016/j.physb.2015.01.036},
url = {https://doi.org/10.1016/j.physb.2015.01.036}
}

@article{Yamaguchi2020AsymmetricJ1J2,
  title = {Variety of order-by-disorder phases in the asymmetric ${J}_{1}\ensuremath{-}{J}_{2}$ zigzag ladder: From the delta chain to the ${J}_{1}\ensuremath{-}{J}_{2}$ chain},
  author = {Yamaguchi, Tomoki and Drechsler, Stefan-Ludwig and Ohta, Yukinori and Nishimoto, Satoshi},
  journal = {Phys. Rev. B},
  volume = {101},
  issue = {10},
  pages = {104407},
  numpages = {18},
  year = {2020},
  month = {Mar},
  publisher = {American Physical Society},
  doi = {10.1103/PhysRevB.101.104407},
  url = {https://link.aps.org/doi/10.1103/PhysRevB.101.104407}
}

@article{Monti1991DeltaTrees,
  title={Spin-12 Heisenberg model on $\Delta$ trees},
  author={Monti, F and S{\"u}to, A},
  journal={Physics Letters A},
  volume = {156},
  number = {3},
  pages = {197-200},
  year = {1991},
  issn = {0375-9601},
  doi = {https://doi.org/10.1016/0375-9601(91)90937-4},
  url = {https://www.sciencedirect.com/science/article/pii/0375960191909374},
}

@article{RuizPerez2000Malonate,
author = {Ruiz-P{'e}rez, Catalina and Hern{'a}ndez-Molina, Mar{'i}a and Lorenzo-Luis, Pablo and Lloret, Francesc and Julve, Miguel and Cano, Juan and others},
title = {Magnetic Coupling through the Carbon Skeleton of Malonate in Two Polymorphs of {[Cu(bpy)(H2O)][Cu(bpy)(mal)(H2O)]}(ClO4)2 (H2mal = malonic acid; bpy = 2,2'-bipyridine)},
journal = {Inorganic Chemistry},
volume = {39},
number = {17},
pages = {3845--3852},
year = {2000},
doi = {10.1021/ic000314n},
url = {https://doi.org/10.1021/ic000314n}
}

@article{TonegawaKaburagi2004DeltaChain,
title={Ground-state properties of an S= 12$\Delta$-chain with ferro-and antiferromagnetic interactions},
journal = {Journal of Magnetism and Magnetic Materials},
volume = {272-276},
pages = {898-899},
year = {2004},
note = {Proceedings of the International Conference on Magnetism (ICM 2003)},
issn = {0304-8853},
doi = {https://doi.org/10.1016/j.jmmm.2003.11.367},
url = {https://www.sciencedirect.com/science/article/pii/S0304885303013027},
author = {Takashi Tonegawa and Makoto Kaburagi},
}

@article{Kaburagi2005AnisotropicDelta,
    author = {Kaburagi, M. and Tonegawa, T. and Kang, M.},
    title={Ground state phase diagrams of an anisotropic spin-12 $\Delta$-chain with ferro-and antiferromagnetic interactions},
    journal = {Journal of Applied Physics},
    volume = {97},
    number = {10},
    pages = {10B306},
    year = {2005},
    month = {05},
    issn = {0021-8979},
    doi = {10.1063/1.1851893},
    url = {https://doi.org/10.1063/1.1851893},
}

@article{Inagaki2005DeltaChainHF,
author = {Inagaki, Yuji and Narumi, Yasuo and Kindo, Koichi and Kikuchi, Hikomitsu and Kamikawa, Tomohisa and Kunimoto, Takashi and Okubo, Susumu and Ohta, Hitoshi and Saito, Takashi and Azuma, Masaki and Takano, Mikio and Nojiri, Hiroyuki and Kaburagi, Makoto and Tonegawa, Takashi},
title = {Ferro-Antiferromagnetic Delta-Chain System Studied by High Field Magnetization Measurements},
journal = {Journal of the Physical Society of Japan},
volume = {74},
number = {10},
pages = {2831--2835},
year = {2005},
doi = {10.1143/JPSJ.74.2831},
url = {https://doi.org/10.1143/JPSJ.74.2831}
}

@article{Blundell2003QuantumTopological,
author = {Blundell, S. A. and N{'u}{~n}ez-Regueiro, M. D.},
title = {Quantum topological excitations: from the sawtooth lattice to the Heisenberg chain},
journal = {The European Physical Journal B},
volume = {31},
number = {4},
pages = {453--456},
year = {2003},
doi = {10.1140/epjb/e2003-00054-2},
url = {https://doi.org/10.1140/epjb/e2003-00054-2}
}

@article{ITensor,
title={{The ITensor Software Library for Tensor Network Calculations}},
author={Fishman, Matthew and White, Steven R. and Stoudenmire, E. Miles},
journal={SciPost Phys. Codebases},
pages={4},
year={2022},
doi={10.21468/SciPostPhysCodeb.4},
url={https://scipost.org/10.21468/SciPostPhysCodeb.4}
}

@article{kingma2015adam,
title = {Adam: A Method for Stochastic Optimization},
author = {Kingma, Diederik P. and Ba, Jimmy},
journal = {International Conference on Learning Representations (ICLR)},
year = {2015},
eprint = {1412.6980},
archivePrefix = {arXiv},
primaryClass = {cs.LG},
url = {https://arxiv.org/abs/1412.6980}
}

@inproceedings{Glorot2010,
  title = {Understanding the difficulty of training deep feedforward neural networks},
  author = {Glorot, Xavier and Bengio, Yoshua},
  booktitle = {Proceedings of the Thirteenth International Conference on Artificial Intelligence and Statistics},
  pages = {249--256},
  year = {2010},
  organization = {PMLR},
  url = {https://proceedings.mlr.press/v9/glorot10a.html}
}

@inproceedings{He2015,
  title = {Delving deep into rectifiers: {S}urpassing human-level performance on {I}mage{N}et classification},
  author = {He, Kaiming and Zhang, Xiangyu and Ren, Shaoqing and Sun, Jian},
  booktitle = {2015 IEEE International Conference on Computer Vision (ICCV)},
  pages = {1026--1034},
  year = {2015},
  organization = {IEEE},
  doi = {10.1109/ICCV.2015.123}
}

@article{Pascanu2012,
  title = {On the difficulty of training {R}ecurrent {N}eural {N}etworks},
  author = {Pascanu, Razvan and Mikolov, Tomas and Bengio, Yoshua},
  journal = {arXiv preprint arXiv:1211.5063},
  year = {2012},
  url = {https://arxiv.org/abs/1211.5063}
}

@inproceedings{Singh2020,
  title = {Cooperative {I}nitialization based {D}eep {N}eural {N}etwork {T}raining},
  author = {Singh, Pankaj and Varshney, Megha and Namboodiri, Vinay P.},
  booktitle = {2020 IEEE Winter Conference on Applications of Computer Vision (WACV)},
  pages = {1130--1139},
  year = {2020},
  organization = {IEEE},
  doi = {10.1109/WACV45572.2020.9093378}
}

@article{rc31-5hl9,
  title = {Simulating the Two-Dimensional $t\text{\ensuremath{-}}J$ Model at Finite Doping with Neural Quantum States},
  author = {Lange, Hannah and B\"ohler, Annika and Roth, Christopher and Bohrdt, Annabelle},
  journal = {Phys. Rev. Lett.},
  volume = {135},
  issue = {13},
  pages = {136504},
  numpages = {7},
  year = {2025},
  month = {Sep},
  publisher = {American Physical Society},
  doi = {10.1103/rc31-5hl9},
  url = {https://link.aps.org/doi/10.1103/rc31-5hl9}
}

@Article{netket3:2022,
    title={NetKet 3: Machine Learning Toolbox for Many-Body Quantum Systems},
    author={Filippo Vicentini and Damian Hofmann and Attila Szabó and Dian Wu and Christopher Roth and Clemens Giuliani and Gabriel Pescia and Jannes Nys and Vladimir Vargas-Calderón and Nikita Astrakhantsev and Giuseppe Carleo},
    journal={SciPost Phys. Codebases},
    pages={7},
    year={2022},
    publisher={SciPost},
    doi={10.21468/SciPostPhysCodeb.7},
    url={https://scipost.org/10.21468/SciPostPhysCodeb.7}
}

@article{netket2:2019,
    title={NetKet: A machine learning toolkit for many-body quantum systems},
    author={Carleo, Giuseppe and Choo, Kenny and Hofmann, Damian and Smith, James ET and Westerhout, Tom and Alet, Fabien and Davis, Emily J and Efthymiou, Stavros and Glasser, Ivan and Lin, Sheng-Hsuan and Mauri, Marta and Mazzola, Guglielmo and Pereira, Christian B and Vicentini, Filippo},
    journal={SoftwareX},
    volume={10},
    pages={100311},
    year={2019},
    publisher={Elsevier},
    doi={10.1016/j.softx.2019.100311},
    url={https://www.sciencedirect.com/science/article/pii/S2352711019300974}
}

@article{ybgv-35jm,
  title = {Design principles of deep translationally symmetric neural quantum states for frustrated magnets},
  author = {Nutakki, Rajah P. and Shokry, Ahmedeo and Vicentini, Filippo},
  journal = {Phys. Rev. Res.},
  volume = {7},
  issue = {4},
  pages = {043099},
  numpages = {13},
  year = {2025},
  month = {Oct},
  publisher = {American Physical Society},
  doi = {10.1103/ybgv-35jm},
  url = {https://link.aps.org/doi/10.1103/ybgv-35jm}
}

\end{document}